\documentclass[a4paper,pra,reprint, twocolumn,superscriptaddress]{revtex4}

\usepackage{ulem}
\usepackage{amssymb}
\usepackage{amsmath}
\usepackage{epsfig}
\usepackage{color}
\usepackage{graphics, graphicx}
\usepackage{bbold}
\usepackage{psfrag}
\usepackage{mathcomp}
\usepackage{amsmath}
\usepackage{amssymb}%
\usepackage{mathrsfs}%
\usepackage{subfigure}
\usepackage{verbatim}
\usepackage{xcolor}
\usepackage[colorlinks,citecolor=blue,urlcolor=blue]{hyperref}
\def\cp#1{\mathbf{#1}}

\usepackage{mathrsfs}

\begin{document}

\date{\today}
\title{Quantum droplet in a mixture of Bose-Fermi superfluids}
\author{Jing-Bo Wang}
\affiliation{CAS Key Laboratory of Quantum Information, University of Science and Technology of China, Hefei 230026, China}
\author{Jian-Song Pan}
\affiliation{Department of Physics, National University of Singapore, Singapore 117542}
\author{Xiaoling Cui}
\email{xlcui@iphy.ac.cn}
\affiliation{Beijing National Laboratory for Condensed Matter Physics, Institute of Physics, Chinese Academy of Sciences, Beijing 100190, China}
\affiliation{Songshan Lake Materials Laboratory , Dongguan, Guangdong 523808, China}
\author{Wei Yi}
\email{wyiz@ustc.edu.cn}
\affiliation{CAS Key Laboratory of Quantum Information, University of Science and Technology of China, Hefei 230026, China}
\affiliation{CAS Center For Excellence in Quantum Information and Quantum Physics, Hefei 230026, China}

\begin{abstract}
We study the formation of quantum droplets in the mixture of a single-component Bose-Einstein condensate (BEC), and a two-species Fermi superfluid across a wide Feshbach resonance. With repulsive boson-boson and attractive boson-fermion interactions, we show that quantum droplets can be stabilized by attractive fermion-fermion interactions in the Bardeen-Cooper-Schieffer (BCS) side of the resonance, and can also exist in deep BEC regime under weak boson-fermion interactions.
We map out the phase diagram for stable droplets with respect to the boson-boson and boson-fermion interactions, and discuss the role of different types of quantum fluctuations in the relevant regions of the BCS-BEC crossover.
Our work reveals the impact of fermion pairing on the formation of quantum droplets in Bose-Fermi mixtures, and provides a useful guide for future experiments.
\end{abstract}

\maketitle

\section{Introduction}

The recent experimental observation of self-bound quantum droplets in dipolar~\cite{2016Kadau,2016Ferrier,2016Ferriertwo,2016Schmitt,2016Chomaz,2019Fabian} or binary Bose-Einstein condensates (BECs)~\cite{2018Cabrera_two,2018Cheiney,2018Semeghini,2019Ferioli,2019Burchianti} has stimulated wide research interest~\cite{2016Santos,2018Cikojevi,2018Wenzel,2018Ancilotto,2019Tengstrand,2019Gautam,2019Cikojevi,2019kartashov,2019Li}. Stabilized by a subtle balance between the attractive mean-field interactions and repulsive quantum fluctuations~\cite{2002Bulgac,2015Petrov}, these exotic liquid-like states are a direct manifestation of quantum many-body effects, and open up the avenue of exploring gas-liquid transition in the quantum regime~\cite{2019He}. As a universal binding mechanism for the quantum droplet, Petrov pointed out in his seminal work that~\cite{2015Petrov}, in a binary BEC, the mean-field inter-species attraction, which tends to collapse the BEC, is counteracted by the Lee-Huang-Yang (LHY) repulsion~\cite{1957Lee}. Based on such an understanding, quantum droplets are also predicted to exist at lower dimensions~\cite{2016Baillie,2016Petrov,2018Sekino,2018Petrov,2018Li}, in spin-orbit-coupled BECs~\cite{2019Chiquillo}, at finite temperatures~\cite{2019He,2019Aybar}, and in photonic systems~\cite{2018Wilson,2018Westerberg}.

While most of the early theoretical studies focus on bosonic systems, people have also started to explore the existence of quantum droplets in Bose-Fermi mixtures~\cite{2007Salasnich,2018Cui,2018Adhikari,2019Rakshit,2019Rakshittwo}, where the coexistence of different statistics or distinct quasiparticles play an important role.
For example, self-bound droplets exist in a mixture of BEC and non-interacting fermions~\cite{2019Rakshit,2019Rakshittwo}, with attractive boson-fermion interactions. In such a system, the large inter-species attraction is balanced by multiple repulsive contributions: the Fermi degenerate pressure, the LHY correction of the BEC, and the second-order correction to the boson-fermion interaction due to density fluctuations~\cite{2002Albus,2002Viverit}. However, owing to the large contribution from Fermi pressure in the dilute limit, the stabilization of droplets requires a strong attractive boson-fermion interaction~\cite{2019Rakshit}, which is unfavorable experimentally, due to the inevitable atom loss in the strongly interacting regime.
As a solution, it has been suggested that, by introducing synthetic spin-orbit coupling to modify low-energy, single-particle dispersion of the non-interacting Fermi gas~\cite{2018Cui}, the Fermi pressure can have more favorable scaling with the fermion density, such that droplets are stabilized under weaker boson-fermion interactions.

Alternatively, the Fermi pressure can be lowered by the formation of Cooper pairs. In particular, assuming the two-component fermions are close to a wide $s$-wave Feshbach resonance, the decrease in Fermi pressure becomes even more pronounced as the system is tuned away from the BCS limit toward the resonance. The fate of droplets in such a system therefore sensitively depends on the interplay of the reduced Fermi pressure and quantum fluctuations, both of which can be quite different from those with a non-interacting Fermi gas. A key question is whether these competing factors inherent in the system can lead to experimentally favorable conditions for the observation of droplets in Bose-Fermi mixtures.

In this work, we study the stability of quantum droplets in a mixture of BEC and Fermi pairing superfluids~\cite{2014Ferrier,2017DeSalvo}, where the fermion-fermion interaction is tuned across a wide $s$-wave Feshbach resonance~\cite{2008Bloch}. Given a fixed repulsive boson-boson interaction strength, we show that the critical boson-fermion interaction strength to support quantum droplets can be further reduced by tuning the fermion-fermion interaction from  Bardeen-Cooper-Schieffer (BCS) side toward resonance. This is facilitated by a modified boson-fermion fluctuation energy, which can turn from positive to attractive as the fermion-fermion interaction is tuned.
This additional attractive contribution to the interaction energy relaxes the requirement on a large boson-fermion interaction, giving rise to a larger stability region of droplets when the system is tuned away from the weak-coupling BCS limit.
On the BEC side, the system can be considered as a mixture of a single-component BEC and a composite BEC of fermion dimers, provided the system is in the deep BEC regime. We show that droplets can be stabilized by the LHY corrections of the two BECs, against the attractive boson-dimer interactions. Remarkably, droplets can be stabilized over a considerable region on the BEC side, and at sufficiently small boson-fermion interaction strengths which are readily achievable in current experiments. Our finding provides a comprehensive understanding of droplet formation in the mixed Bose-Fermi superfluids, and serves as a useful guide for future experiments.

Our work is organized as follows. In Sec.~II, we present our models on the BCS and BEC side of the resonance, respectively. We analyze our main results in Sec.~III, and map out the phase diagram for stable droplets in Sec.~IV. Finally, we summarize in Sec.~V.

\section{Model}
We consider a mixture of a single-component BEC and a two-component Fermi superfluid. The intra-species interaction is repulsive for the BEC, while the inter-species interactions between the bosons and fermions are attractive. The interaction between the two spin species of the Fermi gas can be turned across a wide $s$-wave Feshbach resonance. The Hamiltonian of the system can be written as
\begin{align}
H=&\sum_{\cp k} \epsilon^{b}_{\cp k} b_{k}^{\dagger} b_{k}+ \frac{U_{b}}{2V}\sum_{\cp k,\cp k',\cp q} b_{\cp k}^{\dagger} b_{\cp q-\cp k}^{\dagger} b_{\cp q-\cp k'} b_{\cp k'}\nonumber\\
&+\sum_{\cp k,\sigma}\epsilon^f_{\cp k} f_{k, \sigma}^{\dagger} f_{k, \sigma}+\frac{U_{f}}{V} \sum_{\cp k, \cp k',\cp q} f_{\cp k, \uparrow}^{\dagger} f_{\cp q-\cp k, \downarrow}^{\dagger} f_{\cp q-k', \downarrow} f_{\cp k', \uparrow}\nonumber\\
&+\frac{U_{b f}}{V}\sum_{\cp k,\cp k',\cp q,\sigma}  b_{\cp q-\cp k}^{\dagger} f_{\cp k, \sigma}^{\dagger} f_{\cp q-k', \sigma} b_{\cp k'},
\end{align}
where $\epsilon^{b,f}_{k}=k^2/2m_{b,f}$, $m_b$ ($m_f$) is the mass of boson (fermion) atoms, $b^{\dag}_{\cp k}$ ($b_{\cp k}$) creates (annihilates) a bosonic atom with momentum $\cp k$, $f^{\dag}_{\cp k,\sigma}$ ($f_{\cp k,\sigma}$) creates (annihilates) a fermionic atom with pseudo-spin $\sigma$ ($\sigma=\uparrow,\downarrow$) and momentum $\cp k$. The bare boson-boson interactions $g_b$, the bare boson-fermion interaction, and the bare fermion-fermion interaction are related to the scattering lengths $a_b$, $a_{bf}$, and $a_f$, respectively, through the renormalization relation $1/U_i=1/g_i-(1/V)\sum_{\cp k}m_i/k^2$ ($i=b,bf,f$), where $g_i=4\pi a_i/m_i$, with $m_{bf}=2m_bm_f/(m_b+m_f)$ and $V$ the quantization volume.
Here we take $\hbar=1$, and assume that the interaction strength between bosons and fermions is independent of fermion species. For a repulsively interacting BEC and attractive boson-fermion interaction, we have $g_b>0$ and $g_{bf}<0$.

\begin{figure}[tbp]
  \centering
  \includegraphics[width=9cm]{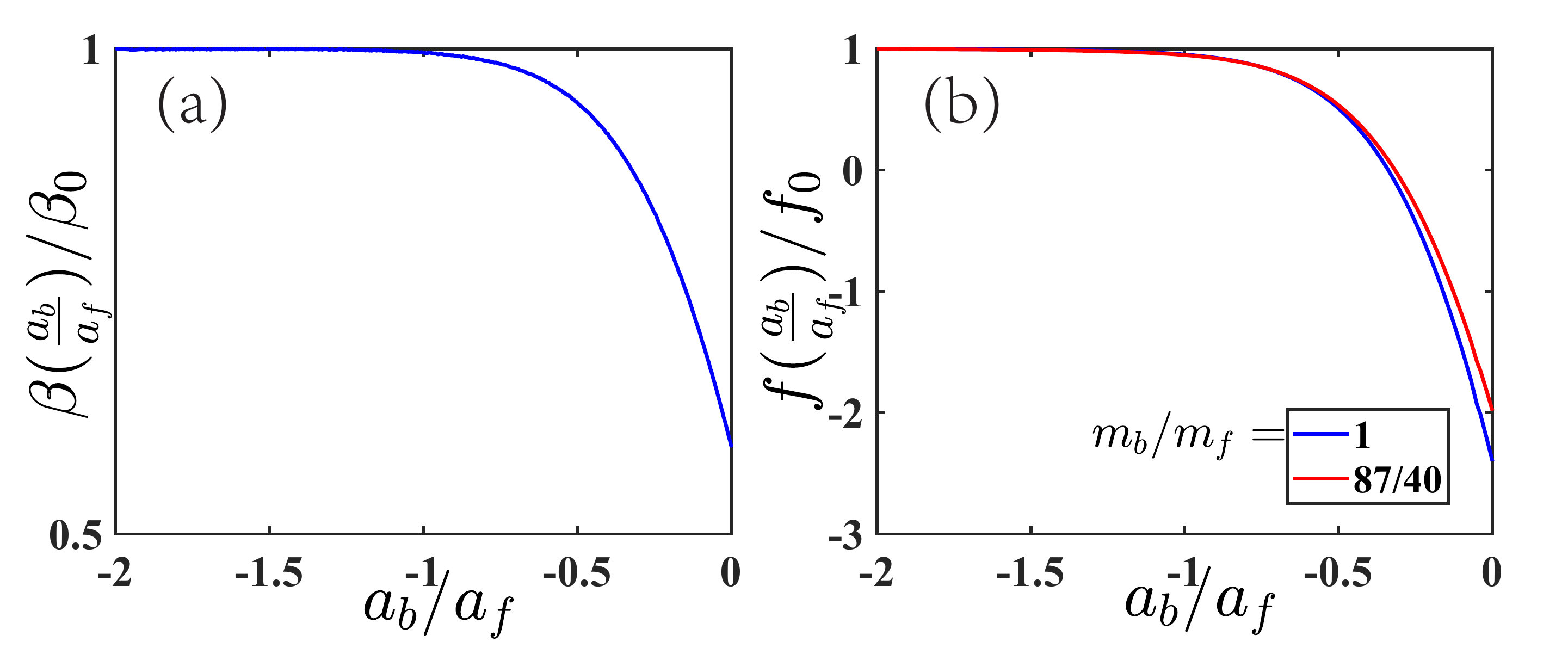}
  \caption{Functions (a) $\beta\left(\frac{a_b}{a_f}\right)$ and (b) $f\left(\frac{m_b}{m_f},\frac{n_b}{n_f},\frac{a_{b}}{a_{f}}\right)$ with increasing $a_b/a_f$ on the BCS side of the resonance. We fix the parameters $a_{bf}/a_b=-10$, and $n_b/n_f=1$. In (b), the blue curve corresponds to the mass ratio $m_b/m_f=1$, and the red curve corresponds to $m_b/m_f=87/40$. Here $\beta_0=3(3\pi^2)^{\frac{2}{3}}/5m_f$ and $f_0=f\left(z,1,\infty\right)$ are respectively proportional to the Fermi pressure and the second-order corrections due to boson-fermion interactions, in the absence of fermion pairing interactions. See Appendix for explicitly expressions for functions $\beta$ and $f$.
  }
  \label{fig:fig1}
\end{figure}

On the BCS side of the Feshbach resonance, the bose-fermion interaction is dominated by collisions of Bogoliubov quasiparticles of the Bose and Fermi superfluids. Assuming small depletions of the superfluids, we rewrite the Hamiltonian as
\begin{align}
H=E_{b}+E_{f}+\sum_{\cp k} \omega_{k} \alpha_{\cp k}^{\dagger} \alpha_{\cp k}+\sum_{\cp k, \sigma} \xi_{k} \beta_{\cp k, \sigma}^{\dagger} \beta_{\cp k, \sigma}+H_{\rm int},
\end{align}
where $\alpha_{\cp k}$ ($\beta_{\cp k,\sigma}$) is the annihilation operator for the boson (fermion) Bogoliubov quasiparticles, with dispersions $\omega_{\cp k}=\sqrt{\epsilon^b_{k}(\epsilon^b_k+2g_bn_b)}$ and $\xi_{\cp k}=\sqrt{(\epsilon^f_k-\mu)^2+\Delta^2}$, respectively. Here $n_b$ is the BEC density, $\mu$ and $\Delta$ are respectively the chemical potential and pairing order parameter of the Fermi superfluid. The ground-state energy densities of the BEC and the Fermi superfluid, defined as $\mathcal{E}_{b,f}=E_{b,f}/V$, are respectively given as
\begin{align}
\mathcal{E}_b&=\frac{g_bn_b^2}{2}+g_{\rm LHY}n_b^{\frac{5}{2}},\label{eq:Eb}\\
\mathcal{E}_f&=\frac{1}{V}\sum_{\cp k} (\epsilon^f_k-\mu-\xi_{\cp k}+\frac{\Delta^2}{2\xi_{\cp k}})+\mu n_f\nonumber\\
&:=\beta\left(\frac{a_b}{a_f}\right) n_{f}^{\frac{5}{3}},\label{eq:Eg}
\end{align}
where the second term in Eq.~(\ref{eq:Eb}) corresponds to the LHY correction, with $g_{\rm LHY}=64 /(15 \sqrt{\pi}) g_{b} a_{b}^{\frac{3 }{2}}$, and $n_f$ is the fermion density.
The function $\beta\left(\frac{a_b}{a_f}\right)$ reflects the reduced Fermi pressure in the presence of pairing superfluid, with $\beta\left(\frac{a_b}{a_f}\right)$ approaching unity in the BCS limit [see Fig.~\ref{fig:fig1}(a) and Appendix].

\begin{figure*}[tbp]
  \includegraphics[width=18cm]{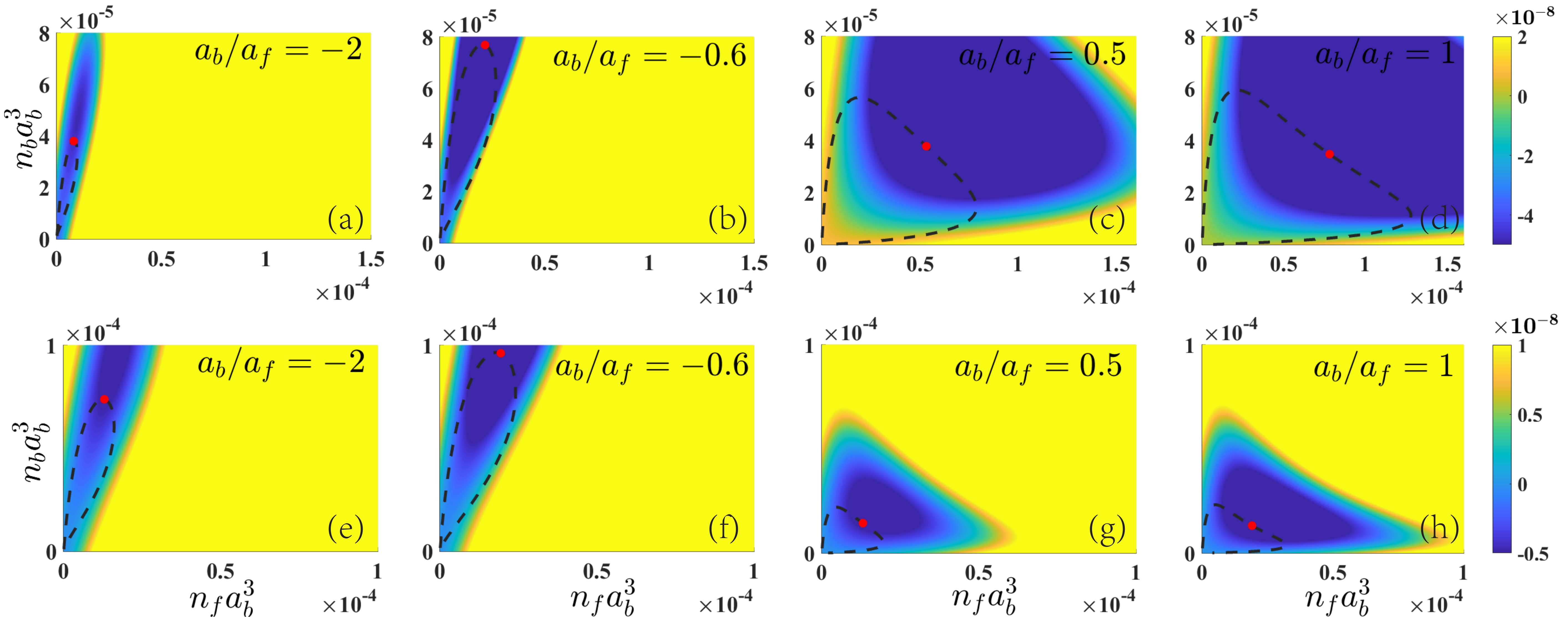}
  \caption{Energy-density contours on the $n_b$--$n_f$ plane, for (a)(b)(c)(d) $m_b/m_f=1$ and (e)(f)(g)(h) $m_b/m_f=87/40$. Contours in (a)(b)(e)(f) are calculated for the BCS side using Eq.~(\ref{eq:energy}), with (a)(e) $a_b/a_f=-2$ and (b)(f) $a_b/a_f=-0.6$.
  Those in (c)(d)(g)(h) are calculated using Eq.~(\ref{eq:BECenergy}) for the BEC region, with (c)(g) $a_b/a_f=0.5$ and (d)(h) $a_b/a_f=1$. The black dashed lines represent zero-pressure ($\mathcal{P}=0$) contours on the density plane and the red dots represent the minimum energy on the contours. }
  \label{fig:fig2}
\end{figure*}

The interaction term can be written in terms of quasiparticle operators as
\begin{align}
&H_{\rm int}=g_{b f} n_{b} \sum_{\cp k, \sigma} l_{\cp k}^{2} \beta_{\cp k, \sigma} \beta_{\cp k, \sigma}^{\dagger}\nonumber\\
&+g_{bf} \sqrt{\frac{n_{b}}{V}} \sum_{\cp k, \cp q, \sigma, \sigma'}\left(u_{\cp k}+v_{\cp k}\right) m_{\cp k+\cp q} l_{\cp q} \alpha_{-\cp k}^{\dagger} \beta_{\cp k+\cp q, \sigma}^\dag \beta_{-\cp q, \sigma'}^{\dagger},\label{eq:eqHint}
\end{align}
where $\{u_{\cp k},v_{\cp k}\}$ ($\{m_{\cp k},l_{\cp k}\}$) are the Bogoliubov coefficients for the boson (fermion) quasiparticles. More explicitly, we have
\begin{align}
u_{\cp k}^2&=1+v_{\cp k}^2=\frac{1}{2}\left(1+\frac{\epsilon^{b}_{k}+g_{b} n_{b}}{\omega_{\cp k}}\right),\\
l_{\cp k}^{2}&=1-m_{\cp k}^2=\frac{1}{2}\left(1-\frac{\epsilon^f_k-\mu}{\xi_{\cp k}}\right).
\end{align}

From Eq.~(\ref{eq:eqHint}), the second-order energy correction for the boson-fermion interaction can be derived, which takes the form
\begin{align}
\mathcal{E}^{(2)}&=g_{b f}^{2} n_f n_{b} \sum_{\cp k} \frac{m_{bf}}{k^{2}}\nonumber\\
&-g_{b f}^{2} n_{b} \sum_{\cp k, \cp q, \sigma, \sigma'} \frac{\left(u_{\cp k}+v_{\cp k}\right)^{2} l^2_{\cp q}m_{\cp q+\cp k}^2}{\omega_{\cp k}+\xi_{\cp k+\cp q, \sigma}+\xi_{\cp q, \sigma'}}\nonumber\\
&-g_{b f}^{2} n_{b} \sum_{\cp k, \cp q, \sigma, \sigma'} \frac{\left(u_{\cp k}+v_{\cp k}\right)^{2} l_{\cp q}m_{\cp q} l_{\cp q+\cp k} m_{\cp q+\cp k}}{\omega_{\cp k}+\xi_{\cp k+\cp q, \sigma}+\xi_{\cp q, \sigma'}}\nonumber\\
&:=g_{b f}^{2} n_{b} n_{f}^{\frac{4}{3}} f\left(\frac{m_b}{m_f},\frac{n_b}{n_f},\frac{a_{b}}{a_{f}}\right),
\end{align}
where the function $f\left(\frac{m_b}{m_f},\frac{n_b}{n_f},\frac{a_{b}}{a_{f}}\right)$ reflects the magnitude of the second-order correction (see Appendix for its explicit form). In Fig.~\ref{fig:fig1}(b), we see that $f\left(\frac{m_b}{m_f},\frac{n_b}{n_f},\frac{a_{b}}{a_{f}}\right)$, being positive in the deep BCS regime, decreases as the fermion-fermion interaction is tuned toward resonance, and becomes negative in the strong-interaction regime on the BCS side of the resonance. Such a behavior signals a qualitative change in the second-order energy correction, from repulsive (in the BCS limit) to attractive (close to resonance), which significantly impacts the droplet formation, as we show later.

It follows from the above derivation that the ground-state energy density of the Bose-Fermi mixture is given by
\begin{align}
\mathcal{E}=&\frac{g_{b} n_{b}^{2}}{2}+\beta\left(\frac{a_b}{a_f}\right) n_{f}^{\frac{5}{ 3}}+g_{b f} n_{b} n_{f}+g_{\rm LHY} n_{b}^{\frac{5}{ 2}}\nonumber\\
&+g_{b f}^{2} n_{b} n_{f}^{\frac{4}{ 3}} f\left(\frac{m_b}{m_f},\frac{n_b}{n_f},\frac{a_{b}}{a_{f}}\right),
\label{eq:energy}
\end{align}
which is applicable on the BCS side of the resonance.

In the deep BEC regime, however, the boson-fermion interaction involves breaking of composite bosons, which are higher-energy processes than the collision between Bose atoms and composite bosons formed by fermions~\cite{2014Zhang,2014Cui}. It is therefore convenient to treat the system as a two-component BEC, with the ground-state energy density given by
\begin{align}
\mathcal{E}=&\frac{g_bn_b^2}{2}+\frac{g_dn_f^2}{2}+g_{bd}n_bn_f\nonumber\\
&+\frac{8}{15\pi^2}m_b^{\frac{3}{2}}(g_bn_b)^{\frac{5}{2}}
g\Big(\frac{2m_f}{m_b},\frac{g_dn_f}{g_bn_b},\frac{g_{bd}^2}{g_dg_b}\Big).
\label{eq:BECenergy}
\end{align}
Here $g_d=4\pi a_d/m_d$,$g_{bd}=2\pi a_{bd}/m_{bd}$, where we take $a_d=0.6a_f$~\cite{2004Petrov}, $a_{bd}=m_{bd}a_{bf}/m_{bf}$~\cite{2014Zhang,2014Cui}, $m_d=2m_f$, $m_{bd}=2m_bm_f/(m_b+2m_f)$. The last term in Eq.~(\ref{eq:BECenergy}) includes LHY corrections for the two-component BEC, with the functional form of $g$ given in the Appendix.

With Eqs.~(\ref{eq:energy}) and (\ref{eq:BECenergy}), we can determine the existence and properties of the ground-state droplets through the following conditions
\begin{align}
\begin{array}{l}{\text { (i) } \mathcal{E}<0, \quad \mathcal{P}=0} \\ {\text { (ii) } \mu_{b} \frac{\partial P}{\partial n_{f}}=\mu_{f} \frac{\partial P}{\partial n_{b}}} \\ {\text { (iii) } \frac{\partial \mu_{b}}{\partial n_{b}}>0, \quad \frac{\partial \mu_{f}}{\partial n_{f}}>0, \quad \frac{\partial \mu_{b}}{\partial n_{b}} \frac{\partial \mu_{f}}{\partial n_{f}}>\left(\frac{\partial \mu_{b}}{\partial n_{f}}\right)^{2}}\end{array},\label{eq:stat}
\end{align}
where $\mu_b=\frac{\partial \mathcal{E}}{\partial n_b}$, $\mu_f=\frac{\partial \mathcal{E}}{\partial n_f}$ and $\mathcal{P}=n_b\mu_b+n_f\mu_f-\mathcal{E}$. We note that the chemical potential $\mu$ and the pairing order parameter $\Delta$ of the Fermi condensate are determined from the standard gap and number equations of the mean-field BCS-BEC crossover theory~\cite{2008Bloch}.

\section{Quantum droplets in mixed Bose-Fermi superfluids}

We are now in a position to numerically study the stability of droplets across the BCS-BEC crossover. However, we emphasize from the outset that, our approach should provides an accurate description in the BCS- or BEC-limit, it should fail in the unitary region close to resonance, due to the sensitivity of droplets to higher-order quantum fluctuations that we neglect in our calculations. In the intermediate region however, our approach should provide a qualitatively valid picture. We therefore mostly focus on the parameter range $|a_b/a_f|>0.5$ in the following.

In Fig.~\ref{fig:fig2}, we show the calculated energy-density contours on the plane of the boson and fermion densities ($n_f$ and $n_b$), with a large, attractive boson-fermion interaction ($a_{bf}/a_b=-10$) on either side of the resonance. Note that we use $a_b$ and $2\pi/(m_b a_b^5)$ as the units for length and energy density, respectively. We see that, regardless of the mass ratio, the energy densities feature a minimum with $\mathcal{E}<0$ in either the deep BCS [Fig.~\ref{fig:fig2}(a)(e)] or the deep BEC regime [Fig.~\ref{fig:fig2}(d)(h)]. However, as the fermion-fermion interaction is tuned toward the resonance, the energy-density minimum moves toward larger densities on the BCS side [Fig.~\ref{fig:fig2}(b)(f)], but toward smaller densities on the BEC side [Fig.~\ref{fig:fig2}(c)(g)]. While the dominance of attractive (repulsive) energy contributions lead to larger (smaller) densities, the tendencies discussed above reflect the different nature and behavior of quantum fluctuations in different regions of the BCS-BEC crossover.

\begin{figure}[tbp]
  \centering
  \includegraphics[width=9cm]{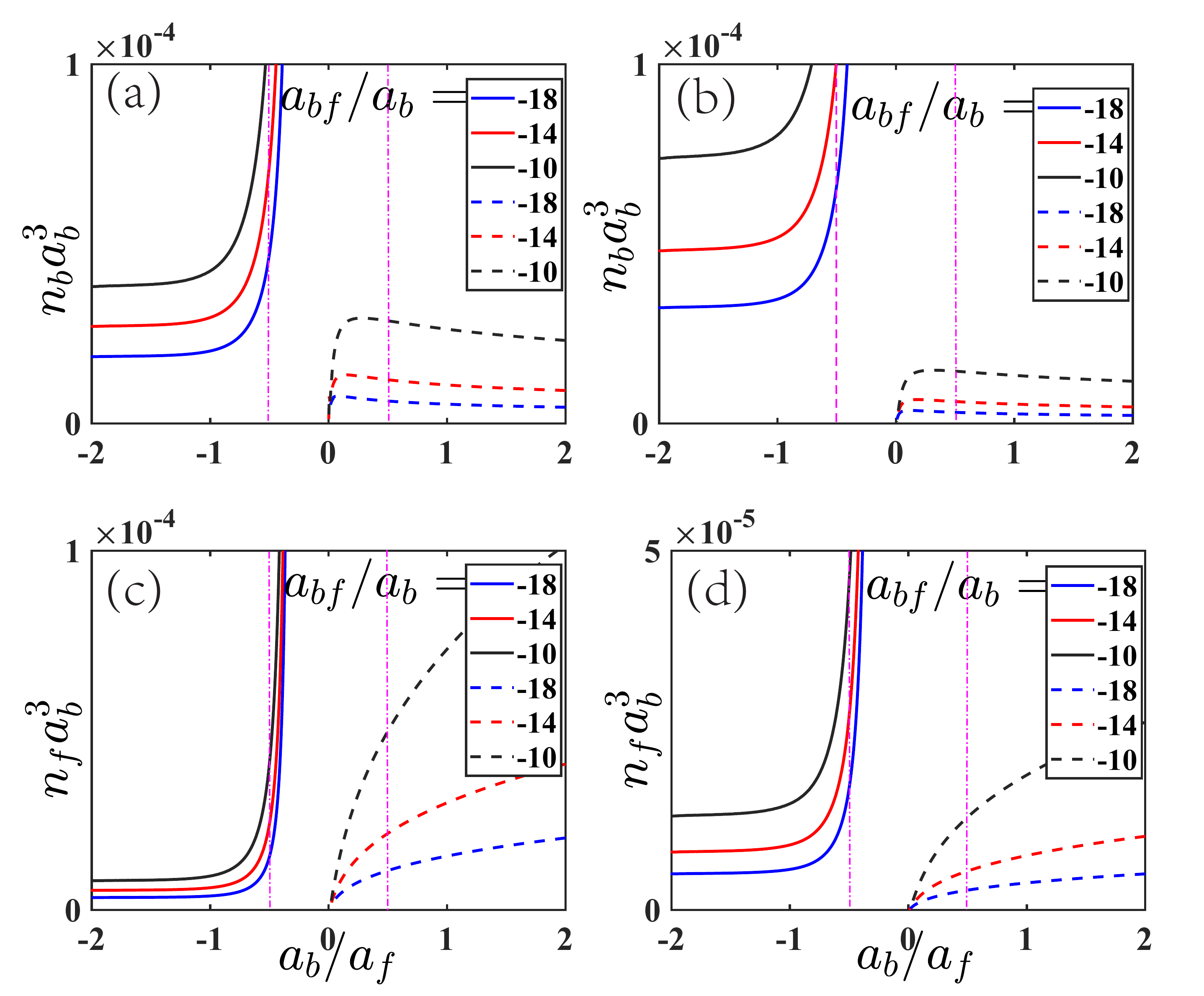}
  \caption{Densities of stable droplets for (a)(c) $m_b/m_f=1$, and (b)(d) $m_b/m_f=87/40$, throughout the BCS-BEC crossover. The blue, red, and black lines respectively correspond to $a_{bf}/a_b=-18$, $a_{bf}/a_b=-14$, and $a_{bf}/a_b=-10$. The solid lines are calculated using Eq.~(\ref{eq:energy}), and the dashed lines are calculated using Eq.~(\ref{eq:BECenergy}). The magenta dash-dotted lines indicate positions of $a_b/a_f=\pm 0.5$.
 }
  \label{fig:fig3}
\end{figure}

To confirm the tendency observed above, we apply the conditions in Eq.~(\ref{eq:stat}) to solve for stable droplets. The resulting zero-pressure contours are shown as black dashed lines in Fig.~\ref{fig:fig2}, against the energy-density contours.
The red dots mark the energy minima along the corresponding contours, which determine the condition for stable droplets. Consistent with the observations above, we find that when the fermion-fermion interaction $a_b/a_f$ is tuned toward resonance, stable droplets on the BCS side feature increased densities, whereas those on the BEC side have decreasing fermion densities but slightly increased boson densities. However, the boson density cannot increase indefinitely along the contour. As the fermion-fermion interaction is tuned even closer to resonance (on the BEC side), the boson density should increase toward the peak value along the zero-pressure contour, before rapidly decreasing on the other side of the peak.

\begin{figure}[tbp]
  \centering
  \includegraphics[width=9cm]{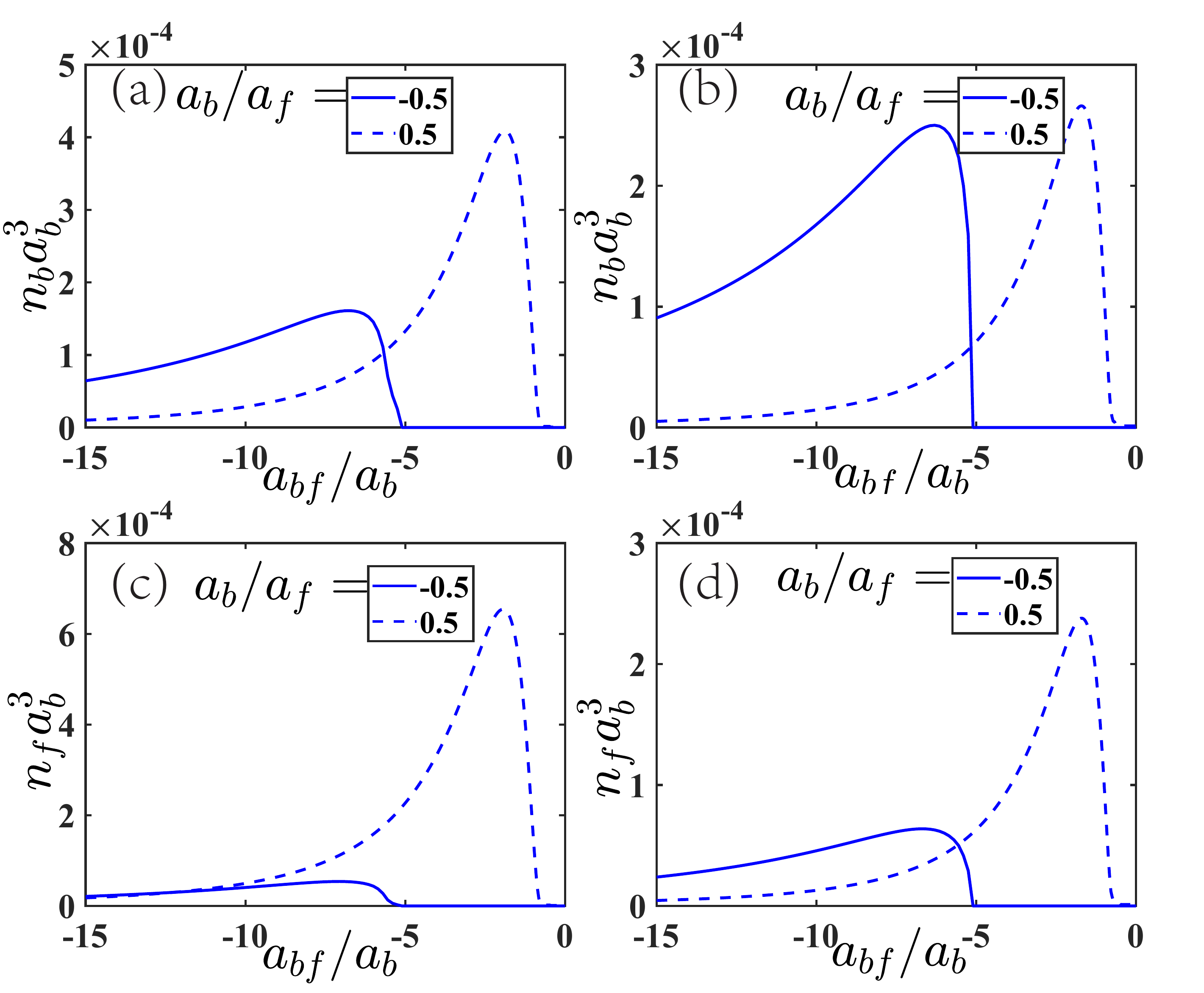}
  \caption{Boson and fermion densities of stable droplets with varying boson-fermi interactions $a_{bf}/a_b$, for (a)(c) $m_b/m_f=1$ and (b)(d) $m_b/m_f=87/40$, respectively. Both the BCS side $a_b/a_f=-0.5$ and BEC side $a_b/a_f=0.5$ are shown.
   the solid lines are calculated using Eq.~(\ref{eq:energy}), and the dashed lines are calculated using Eq.~(\ref{eq:BECenergy}).}
   \label{fig:fig4}
\end{figure}

In Fig.~\ref{fig:fig3}, we explicitly plot the boson and fermion densities for the ground-state droplets as the fermion-fermion interaction is tuned. For both mass ratios $m_b/m_f=1$ [Fig.~\ref{fig:fig3}(a)(c)] and $m_b/m_f=87/40$ [Fig.~\ref{fig:fig3}(b)(d)], the boson and fermion densities monotonically increase toward resonance on the BCS side. In the BCS regime, such an increased density suggests the stabilization of droplets when tuned toward resonance. However, considering atom-loss processes at large densities, the droplets are practically unstable very close to the resonance, once the densities become appreciable.
By contrast, on the BEC side, the fermion density still decreases monotonically toward resonance, while the boson density first increases, but undergoes a rapid drop very close to resonance. We note that such a rapid drop in the boson density on the BEC side could be avoided when fermion-fermion fluctuations are taken into account, which should be attractive~\cite{2015Yi}, and significant close to resonance. Our results thus suggest that, while droplets are stabilized over a considerable parameter region on either side of the resonance, their stability near the resonance sensitively depends on higher-order fluctuations. Furthermore, the stabilization of droplets on different sides of the resonance are due to the quantum fluctuations of different physical origin.
On the BCS side, the dominant fluctuation energy comes from the second-order correction in the boson-fermion interaction which becomes less repulsive and even negative close to resonance. Whereas in the deep BEC regime, it main comes from the repulsive LHY corrections.

On the other hand, under a fixed fermion-fermion interaction on either side of the resonance, both fermion and boson densities of a stable droplet first increase then decrease as the attractive boson-fermion interaction becomes weaker (smaller $|a_{bf}|$). Such a non-monotonic behavior is illustrated in Fig.~\ref{fig:fig4}, which suggests that droplets are destabilized by dominant repulsive energy contributions for sufficiently small boson-fermion interactions. While a similar situation occurs for a Bose-Fermi mixture with non-interacting Fermi gas, a key question here is whether droplets in a mixture of Bose-Fermi mixture can still survive under a weak boson-fermion interaction.

\begin{figure}[tbp]
  \centering
  \includegraphics[width=9cm]{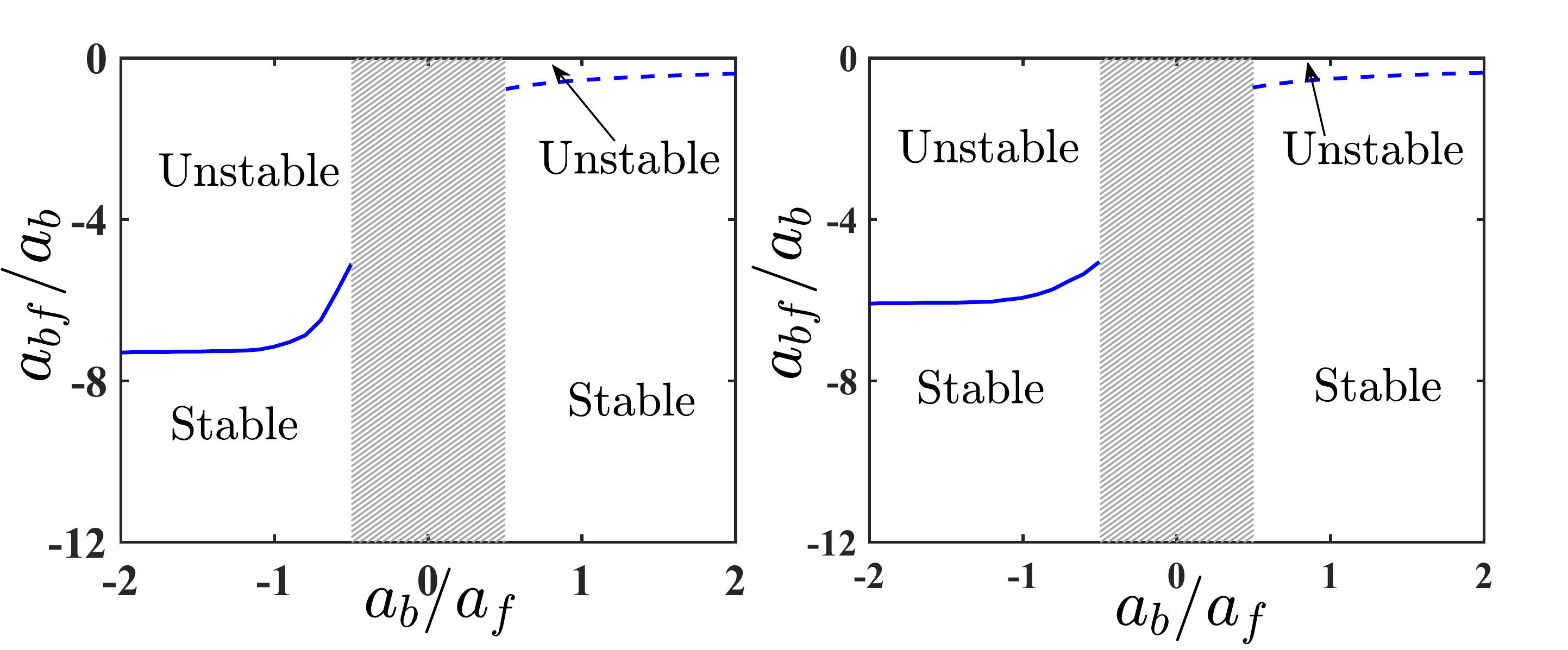}
  \caption{Phase diagram for stable droplets for the mass ratio (a) $m_b/m_f=1$, and (b) $m_b/m_f=87/40$. The shaded area indicate the region $|a_b/a_f|<0.5$, where our approach becomes unreliable. The phase boundaries (solid and dashed blue curves) are determined by the threshold max$(n_ba_b^3,n_fa_b^3)=10^{-7}$. The solid boundary is calculated using Eq.~(\ref{eq:energy}), and the dashed one is calculated using Eq.~(\ref{eq:BECenergy}).}
   \label{fig:fig5}
\end{figure}

\section{Phase diagram}

To address the question, in Fig.~\ref{fig:fig5}, we show the phase diagrams for stable droplets on the $a_{bf}/a_b$--$a_b/a_f$ plane for different mass ratios. As discussed previously, we only focus on the region with $|a_b/a_f|>0.5$ where our approach is qualitatively reliable.
The phase boundaries between the liquid (Stable) and gas (Unstable) regions are determined by the small-density threshold max$(n_ba_b^3,n_fa_b^3)=10^{-7}$ (solid blue lines on the BCS side and dashed blue lines on the BEC side), where densities for stable droplets become vanishingly small under a dominant repulsive energy contribution.

On the BCS side, our phase diagrams clearly indicate the increased stability of droplets at weaker boson-fermion interactions, when the fermions are tuned closer to resonance. This is a direct consequence of the decrease in the repulsive second-order boson-fermion fluctuation energy, as shown in Fig.~\ref{fig:fig1}(b).
Furthermore, the critical boson-fermion interaction $a_{bf}/a_b$ for stable droplets remains small over a large region on the BEC side. For instance, in the BEC limit, the droplets are stable for $a_{bf}/a_b<-0.45$ with $m_b/m_f=1$, and $a_{bf}/a_b<-0.41$ with $m_b/m_f=87/40$.
Thus, an important message here is that stable droplets can be observed in a mixture of Bose-Fermi superfluids under a relatively weak boson-fermion interactions, either on the BCS side or, more dramatically, on the BEC side of the fermion-fermion Feshbach resonance.
As a concrete example, we consider a mixture of $^{87}$Rb and $^{40}$K, close to the wide Feshbach resonance of $^{40}$K near $202$G. Here the scattering lengths are: $95a_0$ between $^{87}$Rb and $^{87}$Rb atoms, $-261a_0$ between $^{87}$Rb and $^{40}$K, with $a_0$ the Bohr radius. We therefore have $a_{bf}/a_b\approx-2.75$, which should feature stable droplets on the BEC side of the resonance, according to the phase diagram. Further, for a typical density $n_b=5\times10^{-14}\text{cm}^{-3}$, we have $n_ba_b^3\approx 10^{-5}$, which is on the order of stable densities shown in Fig.~\ref{fig:fig3}.

\section{Summary and discussion}
To summarize, we study the formation of quantum droplets in a mixture of Bose-Fermi superfluids, where the fermion-fermion interaction is tuned  across a wide Feshbach resonance. We show that droplets can be stabilized over a considerable region on both sides of the Feshbach resonance, facilitated by fluctuations associated with fermion pairing.
Our key finding is that it is possible to stabilize droplets under a weak boson-fermion interaction, particularly on the BEC side of the fermion-fermion Feshbach resonance. This should facilitate the experimental observation of droplets in Bose-Fermi mixtures, where
experimentally attainable $a_{bf}$ is typically small in the absence of a boson-fermion Feshbach resonance.
Finally, we note that we have neglected fluctuation energy of the fermion-fermion interaction for our calculation, as well as higher-order fluctuations. These contributions should ultimately determine the fate of droplets in the unitary region close to the Feshbach resonance. For example, the inclusion of the attractive, second-order correction of the fermion-fermion fluctuation~\cite{2015Yi} would provide a competition against the LHY corrections on the BEC side, thus further enhance droplet formation on the BEC side. However, the same contribution would
further increase the density of stable droplets close to resonance on the BCS side, which would practically destabilize the system due to atom loss. These considerations suggest that droplet formation in the unitary region is still an interesting open question, which we leave to future studies.

\section*{Acknowledgements}
We thank Zengqiang Yu for helpful comments and discussions. This work has been supported by the Natural Science Foundation of China (11974331, 11421092, 11534014) and the National Key R\&D Program (Grant Nos. 2016YFA0301700, 2017YFA0304100, 2018YFA0307600, 2016YFA0300603).

\begin{widetext}
\appendix

\section{Definition of various functions}

In this Appendix, we define the dimensionless functions $\beta(\frac{a_b}{a_f})$, $g\Big(\frac{2m_f}{m_b},\frac{g_dn_f}{g_bn_b},\frac{g_{bd}^2}{g_dg_b}\Big)$, and $f\left(\frac{m_b}{m_f},\frac{n_b}{n_f},\frac{a_{b}}{a_{f}}\right)$. Explicitly, we have
\begin{align}
&\beta\left(\frac{a_b}{a_f}\right)=\frac{\left(3 \pi^{2}\right)^{\frac{2}{3}}}{2 m_{f}} \Big\{\frac{3}{2}\int_{0}^{\infty} t^{2} d t
\left[t^{2}-\tilde{\mu}-\sqrt{\left(t^{2}-\tilde{\mu}\right)^{2}+\tilde{\Delta}^{2}}+\frac{\tilde{\Delta}^{2}}{2 \sqrt{\left(t^{2}-\tilde{\mu}\right)^{2}+\tilde{\Delta}^{2}}}\right]+\tilde\mu\Big\},\\
&g(z,y,x)=\frac{15}{32}\int_0^\infty t^2dt\left[\sum_{\pm}\sqrt{\frac{\mathcal{G_{+}}}{2} \pm \sqrt{\frac{\mathcal{G}^2_{-}}{4}+\frac{xt^4}{z}}} -\left(\frac{t^{2}}{2}+1\right)-\left(\frac{t^{2}}{2 z}+y\right)+\frac{1+z y^{2}+4 \frac{z}{1+z} x}{t^{2}}\right],\\
&\mathcal{G}_{\pm}(k,z,y)=\frac{k^2}{2}(\frac{k^2}{2}+2)\pm\frac{k^2}{2}(\frac{k^2}{2z}+2y),\\
&f\left(z, \frac{n_{b}}{n_{f}}, \frac{a_{b}}{a_{f}}\right)=\frac{\left(3 \pi^{2}\right)^{\frac{2}{3}}}{2 \pi^{2}} \int_{0}^{\infty} d t\left\{\frac{1+z}{z}-\frac{3}{2} \int_{0}^{\infty} h^{2} d h \int_{-1}^{1} d \Omega \frac{z t^{3}}{\sqrt{t^{2}+\frac{4 \pi a_{b} n_{b}}{\left(3 \pi^{2} n_{f}\right)^{\frac{2}{3}}}}}\right.\nonumber\\
&\left.\times \frac{\left[1-\frac{t^{2}-\tilde{\mu}}{\sqrt{\left(t^{2}-\tilde{\mu}\right)^{2}+\tilde{\Delta}^{2}}}\right]\left[1+\frac{t^{2}+h^{2}+2 h t \Omega-\tilde{\mu}}{\sqrt{\left(t^{2}+h^{2}+2 h t \Omega-\tilde{\mu}\right)^{2}+\tilde{\Delta}^{2}}}\right]+\frac{\tilde{\Delta}^2}
{\sqrt{(t^2-\tilde{\mu})^2+\tilde{\Delta}^2}\sqrt{(t^2+h^2+2ht\Omega-\hat\mu)^2+\tilde{\Delta}^2}}}{t \sqrt{t^{2}+\frac{4 \pi a_{b} n_{b}}{\left(3 \pi^2 n_{f}\right)^{\frac{2}{3}}}}+\sqrt{\left(t^{2}+h^{2}+2 h t \Omega-\tilde{\mu}\right)^{2}+\tilde{\Delta}^{2}}+\sqrt{\left(t^{2}-\tilde{\mu}\right)^{2}+\tilde{\Delta}^{2}}}\right\}.
\end{align}
Here $\tilde \mu=\mu/\epsilon_F$, $\tilde\Delta=\Delta/\epsilon_F$, with $\epsilon_F=(3\pi^2 n_f)^{\frac{2}{3}}/2m_f$.
\end{widetext}


\bibliographystyle{apsrev4-1}
\bibliography{droplet_reference}

\begin{thebibliography}{47}%
\makeatletter
\providecommand \@ifxundefined [1]{%
 \@ifx{#1\undefined}
}%
\providecommand \@ifnum [1]{%
 \ifnum #1\expandafter \@firstoftwo
 \else \expandafter \@secondoftwo
 \fi
}%
\providecommand \@ifx [1]{%
 \ifx #1\expandafter \@firstoftwo
 \else \expandafter \@secondoftwo
 \fi
}%
\providecommand \natexlab [1]{#1}%
\providecommand \enquote  [1]{``#1''}%
\providecommand \bibnamefont  [1]{#1}%
\providecommand \bibfnamefont [1]{#1}%
\providecommand \citenamefont [1]{#1}%
\providecommand \href@noop [0]{\@secondoftwo}%
\providecommand \href [0]{\begingroup \@sanitize@url \@href}%
\providecommand \@href[1]{\@@startlink{#1}\@@href}%
\providecommand \@@href[1]{\endgroup#1\@@endlink}%
\providecommand \@sanitize@url [0]{\catcode `\\12\catcode `\$12\catcode
  `\&12\catcode `\#12\catcode `\^12\catcode `\_12\catcode `\%12\relax}%
\providecommand \@@startlink[1]{}%
\providecommand \@@endlink[0]{}%
\providecommand \url  [0]{\begingroup\@sanitize@url \@url }%
\providecommand \@url [1]{\endgroup\@href {#1}{\urlprefix }}%
\providecommand \urlprefix  [0]{URL }%
\providecommand \Eprint [0]{\href }%
\providecommand \doibase [0]{http://dx.doi.org/}%
\providecommand \selectlanguage [0]{\@gobble}%
\providecommand \bibinfo  [0]{\@secondoftwo}%
\providecommand \bibfield  [0]{\@secondoftwo}%
\providecommand \translation [1]{[#1]}%
\providecommand \BibitemOpen [0]{}%
\providecommand \bibitemStop [0]{}%
\providecommand \bibitemNoStop [0]{.\EOS\space}%
\providecommand \EOS [0]{\spacefactor3000\relax}%
\providecommand \BibitemShut  [1]{\csname bibitem#1\endcsname}%
\let\auto@bib@innerbib\@empty
\bibitem [{\citenamefont {Kadau}\ \emph {et~al.}(2016)\citenamefont {Kadau},
  \citenamefont {Schmitt}, \citenamefont {Wenzel}, \citenamefont {Wink},
  \citenamefont {Maier}, \citenamefont {Ferrier-Barbut},\ and\ \citenamefont
  {Pfau}}]{2016Kadau}%
  \BibitemOpen
  \bibfield  {author} {\bibinfo {author} {\bibfnamefont {H.}~\bibnamefont
  {Kadau}}, \bibinfo {author} {\bibfnamefont {M.}~\bibnamefont {Schmitt}},
  \bibinfo {author} {\bibfnamefont {M.}~\bibnamefont {Wenzel}}, \bibinfo
  {author} {\bibfnamefont {C.}~\bibnamefont {Wink}}, \bibinfo {author}
  {\bibfnamefont {T.}~\bibnamefont {Maier}}, \bibinfo {author} {\bibfnamefont
  {I.}~\bibnamefont {Ferrier-Barbut}}, \ and\ \bibinfo {author} {\bibfnamefont
  {T.}~\bibnamefont {Pfau}},\ }\href@noop {} {\bibfield  {journal} {\bibinfo
  {journal} {Nature}\ }\textbf {\bibinfo {volume} {530}},\ \bibinfo {pages}
  {194} (\bibinfo {year} {2016})}\BibitemShut {NoStop}%
\bibitem [{\citenamefont {Ferrier-Barbut}\ \emph
  {et~al.}(2016{\natexlab{a}})\citenamefont {Ferrier-Barbut}, \citenamefont
  {Kadau}, \citenamefont {Schmitt}, \citenamefont {Wenzel},\ and\ \citenamefont
  {Pfau}}]{2016Ferrier}%
  \BibitemOpen
  \bibfield  {author} {\bibinfo {author} {\bibfnamefont {I.}~\bibnamefont
  {Ferrier-Barbut}}, \bibinfo {author} {\bibfnamefont {H.}~\bibnamefont
  {Kadau}}, \bibinfo {author} {\bibfnamefont {M.}~\bibnamefont {Schmitt}},
  \bibinfo {author} {\bibfnamefont {M.}~\bibnamefont {Wenzel}}, \ and\ \bibinfo
  {author} {\bibfnamefont {T.}~\bibnamefont {Pfau}},\ }\href {\doibase
  10.1103/PhysRevLett.116.215301} {\bibfield  {journal} {\bibinfo  {journal}
  {Phys. Rev. Lett.}\ }\textbf {\bibinfo {volume} {116}},\ \bibinfo {pages}
  {215301} (\bibinfo {year} {2016}{\natexlab{a}})}\BibitemShut {NoStop}%
\bibitem [{\citenamefont {Ferrier-Barbut}\ \emph
  {et~al.}(2016{\natexlab{b}})\citenamefont {Ferrier-Barbut}, \citenamefont
  {Schmitt}, \citenamefont {Wenzel}, \citenamefont {Kadau},\ and\ \citenamefont
  {Pfau}}]{2016Ferriertwo}%
  \BibitemOpen
  \bibfield  {author} {\bibinfo {author} {\bibfnamefont {I.}~\bibnamefont
  {Ferrier-Barbut}}, \bibinfo {author} {\bibfnamefont {M.}~\bibnamefont
  {Schmitt}}, \bibinfo {author} {\bibfnamefont {M.}~\bibnamefont {Wenzel}},
  \bibinfo {author} {\bibfnamefont {H.}~\bibnamefont {Kadau}}, \ and\ \bibinfo
  {author} {\bibfnamefont {T.}~\bibnamefont {Pfau}},\ }\href@noop {} {\bibfield
   {journal} {\bibinfo  {journal} {J. Phys. B}\ }\textbf {\bibinfo {volume}
  {49}},\ \bibinfo {pages} {214004} (\bibinfo {year}
  {2016}{\natexlab{b}})}\BibitemShut {NoStop}%
\bibitem [{\citenamefont {Schmitt}\ \emph {et~al.}(2016)\citenamefont
  {Schmitt}, \citenamefont {Wenzel}, \citenamefont {B{\"o}ttcher},
  \citenamefont {Ferrier-Barbut},\ and\ \citenamefont {Pfau}}]{2016Schmitt}%
  \BibitemOpen
  \bibfield  {author} {\bibinfo {author} {\bibfnamefont {M.}~\bibnamefont
  {Schmitt}}, \bibinfo {author} {\bibfnamefont {M.}~\bibnamefont {Wenzel}},
  \bibinfo {author} {\bibfnamefont {F.}~\bibnamefont {B{\"o}ttcher}}, \bibinfo
  {author} {\bibfnamefont {I.}~\bibnamefont {Ferrier-Barbut}}, \ and\ \bibinfo
  {author} {\bibfnamefont {T.}~\bibnamefont {Pfau}},\ }\href@noop {} {\bibfield
   {journal} {\bibinfo  {journal} {Nature}\ }\textbf {\bibinfo {volume}
  {539}},\ \bibinfo {pages} {259} (\bibinfo {year} {2016})}\BibitemShut
  {NoStop}%
\bibitem [{\citenamefont {Chomaz}\ \emph {et~al.}(2016)\citenamefont {Chomaz},
  \citenamefont {Baier}, \citenamefont {Petter}, \citenamefont {Mark},
  \citenamefont {W\"achtler}, \citenamefont {Santos},\ and\ \citenamefont
  {Ferlaino}}]{2016Chomaz}%
  \BibitemOpen
  \bibfield  {author} {\bibinfo {author} {\bibfnamefont {L.}~\bibnamefont
  {Chomaz}}, \bibinfo {author} {\bibfnamefont {S.}~\bibnamefont {Baier}},
  \bibinfo {author} {\bibfnamefont {D.}~\bibnamefont {Petter}}, \bibinfo
  {author} {\bibfnamefont {M.~J.}\ \bibnamefont {Mark}}, \bibinfo {author}
  {\bibfnamefont {F.}~\bibnamefont {W\"achtler}}, \bibinfo {author}
  {\bibfnamefont {L.}~\bibnamefont {Santos}}, \ and\ \bibinfo {author}
  {\bibfnamefont {F.}~\bibnamefont {Ferlaino}},\ }\href {\doibase
  10.1103/PhysRevX.6.041039} {\bibfield  {journal} {\bibinfo  {journal} {Phys.
  Rev. X}\ }\textbf {\bibinfo {volume} {6}},\ \bibinfo {pages} {041039}
  (\bibinfo {year} {2016})}\BibitemShut {NoStop}%
\bibitem [{\citenamefont {B\"ottcher}\ \emph {et~al.}(2019)\citenamefont
  {B\"ottcher}, \citenamefont {Schmidt}, \citenamefont {Wenzel}, \citenamefont
  {Hertkorn}, \citenamefont {Guo}, \citenamefont {Langen},\ and\ \citenamefont
  {Pfau}}]{2019Fabian}%
  \BibitemOpen
  \bibfield  {author} {\bibinfo {author} {\bibfnamefont {F.}~\bibnamefont
  {B\"ottcher}}, \bibinfo {author} {\bibfnamefont {J.-N.}\ \bibnamefont
  {Schmidt}}, \bibinfo {author} {\bibfnamefont {M.}~\bibnamefont {Wenzel}},
  \bibinfo {author} {\bibfnamefont {J.}~\bibnamefont {Hertkorn}}, \bibinfo
  {author} {\bibfnamefont {M.}~\bibnamefont {Guo}}, \bibinfo {author}
  {\bibfnamefont {T.}~\bibnamefont {Langen}}, \ and\ \bibinfo {author}
  {\bibfnamefont {T.}~\bibnamefont {Pfau}},\ }\href {\doibase
  10.1103/PhysRevX.9.011051} {\bibfield  {journal} {\bibinfo  {journal} {Phys.
  Rev. X}\ }\textbf {\bibinfo {volume} {9}},\ \bibinfo {pages} {011051}
  (\bibinfo {year} {2019})}\BibitemShut {NoStop}%
\bibitem [{\citenamefont {Cabrera}\ \emph {et~al.}(2018)\citenamefont
  {Cabrera}, \citenamefont {Tanzi}, \citenamefont {Sanz}, \citenamefont
  {Naylor}, \citenamefont {Thomas}, \citenamefont {Cheiney},\ and\
  \citenamefont {Tarruell}}]{2018Cabrera_two}%
  \BibitemOpen
  \bibfield  {author} {\bibinfo {author} {\bibfnamefont {C.}~\bibnamefont
  {Cabrera}}, \bibinfo {author} {\bibfnamefont {L.}~\bibnamefont {Tanzi}},
  \bibinfo {author} {\bibfnamefont {J.}~\bibnamefont {Sanz}}, \bibinfo {author}
  {\bibfnamefont {B.}~\bibnamefont {Naylor}}, \bibinfo {author} {\bibfnamefont
  {P.}~\bibnamefont {Thomas}}, \bibinfo {author} {\bibfnamefont
  {P.}~\bibnamefont {Cheiney}}, \ and\ \bibinfo {author} {\bibfnamefont
  {L.}~\bibnamefont {Tarruell}},\ }\href@noop {} {\bibfield  {journal}
  {\bibinfo  {journal} {Science}\ }\textbf {\bibinfo {volume} {359}},\ \bibinfo
  {pages} {301} (\bibinfo {year} {2018})}\BibitemShut {NoStop}%
\bibitem [{\citenamefont {Cheiney}\ \emph {et~al.}(2018)\citenamefont
  {Cheiney}, \citenamefont {Cabrera}, \citenamefont {Sanz}, \citenamefont
  {Naylor}, \citenamefont {Tanzi},\ and\ \citenamefont
  {Tarruell}}]{2018Cheiney}%
  \BibitemOpen
  \bibfield  {author} {\bibinfo {author} {\bibfnamefont {P.}~\bibnamefont
  {Cheiney}}, \bibinfo {author} {\bibfnamefont {C.~R.}\ \bibnamefont
  {Cabrera}}, \bibinfo {author} {\bibfnamefont {J.}~\bibnamefont {Sanz}},
  \bibinfo {author} {\bibfnamefont {B.}~\bibnamefont {Naylor}}, \bibinfo
  {author} {\bibfnamefont {L.}~\bibnamefont {Tanzi}}, \ and\ \bibinfo {author}
  {\bibfnamefont {L.}~\bibnamefont {Tarruell}},\ }\href {\doibase
  10.1103/PhysRevLett.120.135301} {\bibfield  {journal} {\bibinfo  {journal}
  {Phys. Rev. Lett.}\ }\textbf {\bibinfo {volume} {120}},\ \bibinfo {pages}
  {135301} (\bibinfo {year} {2018})}\BibitemShut {NoStop}%
\bibitem [{\citenamefont {Semeghini}\ \emph {et~al.}(2018)\citenamefont
  {Semeghini}, \citenamefont {Ferioli}, \citenamefont {Masi}, \citenamefont
  {Mazzinghi}, \citenamefont {Wolswijk}, \citenamefont {Minardi}, \citenamefont
  {Modugno}, \citenamefont {Modugno}, \citenamefont {Inguscio},\ and\
  \citenamefont {Fattori}}]{2018Semeghini}%
  \BibitemOpen
  \bibfield  {author} {\bibinfo {author} {\bibfnamefont {G.}~\bibnamefont
  {Semeghini}}, \bibinfo {author} {\bibfnamefont {G.}~\bibnamefont {Ferioli}},
  \bibinfo {author} {\bibfnamefont {L.}~\bibnamefont {Masi}}, \bibinfo {author}
  {\bibfnamefont {C.}~\bibnamefont {Mazzinghi}}, \bibinfo {author}
  {\bibfnamefont {L.}~\bibnamefont {Wolswijk}}, \bibinfo {author}
  {\bibfnamefont {F.}~\bibnamefont {Minardi}}, \bibinfo {author} {\bibfnamefont
  {M.}~\bibnamefont {Modugno}}, \bibinfo {author} {\bibfnamefont
  {G.}~\bibnamefont {Modugno}}, \bibinfo {author} {\bibfnamefont
  {M.}~\bibnamefont {Inguscio}}, \ and\ \bibinfo {author} {\bibfnamefont
  {M.}~\bibnamefont {Fattori}},\ }\href {\doibase
  10.1103/PhysRevLett.120.235301} {\bibfield  {journal} {\bibinfo  {journal}
  {Phys. Rev. Lett.}\ }\textbf {\bibinfo {volume} {120}},\ \bibinfo {pages}
  {235301} (\bibinfo {year} {2018})}\BibitemShut {NoStop}%
\bibitem [{\citenamefont {Ferioli}\ \emph {et~al.}(2019)\citenamefont
  {Ferioli}, \citenamefont {Semeghini}, \citenamefont {Masi}, \citenamefont
  {Giusti}, \citenamefont {Modugno}, \citenamefont {Inguscio}, \citenamefont
  {Gallem\'{\i}}, \citenamefont {Recati},\ and\ \citenamefont
  {Fattori}}]{2019Ferioli}%
  \BibitemOpen
  \bibfield  {author} {\bibinfo {author} {\bibfnamefont {G.}~\bibnamefont
  {Ferioli}}, \bibinfo {author} {\bibfnamefont {G.}~\bibnamefont {Semeghini}},
  \bibinfo {author} {\bibfnamefont {L.}~\bibnamefont {Masi}}, \bibinfo {author}
  {\bibfnamefont {G.}~\bibnamefont {Giusti}}, \bibinfo {author} {\bibfnamefont
  {G.}~\bibnamefont {Modugno}}, \bibinfo {author} {\bibfnamefont
  {M.}~\bibnamefont {Inguscio}}, \bibinfo {author} {\bibfnamefont
  {A.}~\bibnamefont {Gallem\'{\i}}}, \bibinfo {author} {\bibfnamefont
  {A.}~\bibnamefont {Recati}}, \ and\ \bibinfo {author} {\bibfnamefont
  {M.}~\bibnamefont {Fattori}},\ }\href {\doibase
  10.1103/PhysRevLett.122.090401} {\bibfield  {journal} {\bibinfo  {journal}
  {Phys. Rev. Lett.}\ }\textbf {\bibinfo {volume} {122}},\ \bibinfo {pages}
  {090401} (\bibinfo {year} {2019})}\BibitemShut {NoStop}%
\bibitem [{\citenamefont {D'Errico}\ \emph {et~al.}(2019)\citenamefont
  {D'Errico}, \citenamefont {Burchianti}, \citenamefont {Prevedelli},
  \citenamefont {Salasnich}, \citenamefont {Ancilotto}, \citenamefont
  {Modugno}, \citenamefont {Minardi},\ and\ \citenamefont
  {Fort}}]{2019Burchianti}%
  \BibitemOpen
  \bibfield  {author} {\bibinfo {author} {\bibfnamefont {C.}~\bibnamefont
  {D'Errico}}, \bibinfo {author} {\bibfnamefont {A.}~\bibnamefont
  {Burchianti}}, \bibinfo {author} {\bibfnamefont {M.}~\bibnamefont
  {Prevedelli}}, \bibinfo {author} {\bibfnamefont {L.}~\bibnamefont
  {Salasnich}}, \bibinfo {author} {\bibfnamefont {F.}~\bibnamefont
  {Ancilotto}}, \bibinfo {author} {\bibfnamefont {M.}~\bibnamefont {Modugno}},
  \bibinfo {author} {\bibfnamefont {F.}~\bibnamefont {Minardi}}, \ and\
  \bibinfo {author} {\bibfnamefont {C.}~\bibnamefont {Fort}},\ }\href {\doibase
  10.1103/PhysRevResearch.1.033155} {\bibfield  {journal} {\bibinfo  {journal}
  {Phys. Rev. Research}\ }\textbf {\bibinfo {volume} {1}},\ \bibinfo {pages}
  {033155} (\bibinfo {year} {2019})}\BibitemShut {NoStop}%
\bibitem [{\citenamefont {W\"achtler}\ and\ \citenamefont
  {Santos}(2016)}]{2016Santos}%
  \BibitemOpen
  \bibfield  {author} {\bibinfo {author} {\bibfnamefont {F.}~\bibnamefont
  {W\"achtler}}\ and\ \bibinfo {author} {\bibfnamefont {L.}~\bibnamefont
  {Santos}},\ }\href {\doibase 10.1103/PhysRevA.93.061603} {\bibfield
  {journal} {\bibinfo  {journal} {Phys. Rev. A}\ }\textbf {\bibinfo {volume}
  {93}},\ \bibinfo {pages} {061603} (\bibinfo {year} {2016})}\BibitemShut
  {NoStop}%
\bibitem [{\citenamefont {Cikojevi\ifmmode~\acute{c}\else \'{c}\fi{}}\ \emph
  {et~al.}(2018)\citenamefont {Cikojevi\ifmmode~\acute{c}\else \'{c}\fi{}},
  \citenamefont {D\ifmmode~\check{z}\else \v{z}\fi{}elalija}, \citenamefont
  {Stipanovi\ifmmode~\acute{c}\else \'{c}\fi{}}, \citenamefont {Vranje\ifmmode
  \check{s}\else \v{s}\fi{} Marki\ifmmode~\acute{c}\else \'{c}\fi{}},\ and\
  \citenamefont {Boronat}}]{2018Cikojevi}%
  \BibitemOpen
  \bibfield  {author} {\bibinfo {author} {\bibfnamefont {V.}~\bibnamefont
  {Cikojevi\ifmmode~\acute{c}\else \'{c}\fi{}}}, \bibinfo {author}
  {\bibfnamefont {K.}~\bibnamefont {D\ifmmode~\check{z}\else
  \v{z}\fi{}elalija}}, \bibinfo {author} {\bibfnamefont {P.}~\bibnamefont
  {Stipanovi\ifmmode~\acute{c}\else \'{c}\fi{}}}, \bibinfo {author}
  {\bibfnamefont {L.}~\bibnamefont {Vranje\ifmmode \check{s}\else \v{s}\fi{}
  Marki\ifmmode~\acute{c}\else \'{c}\fi{}}}, \ and\ \bibinfo {author}
  {\bibfnamefont {J.}~\bibnamefont {Boronat}},\ }\href {\doibase
  10.1103/PhysRevB.97.140502} {\bibfield  {journal} {\bibinfo  {journal} {Phys.
  Rev. B}\ }\textbf {\bibinfo {volume} {97}},\ \bibinfo {pages} {140502}
  (\bibinfo {year} {2018})}\BibitemShut {NoStop}%
\bibitem [{\citenamefont {Wenzel}\ \emph {et~al.}(2018)\citenamefont {Wenzel},
  \citenamefont {Pfau},\ and\ \citenamefont {Ferrier-Barbut}}]{2018Wenzel}%
  \BibitemOpen
  \bibfield  {author} {\bibinfo {author} {\bibfnamefont {M.}~\bibnamefont
  {Wenzel}}, \bibinfo {author} {\bibfnamefont {T.}~\bibnamefont {Pfau}}, \ and\
  \bibinfo {author} {\bibfnamefont {I.}~\bibnamefont {Ferrier-Barbut}},\ }\href
  {\doibase 10.1088/1402-4896/aadd72} {\bibfield  {journal} {\bibinfo
  {journal} {Phys. Scr}\ }\textbf {\bibinfo {volume} {93}},\ \bibinfo {pages}
  {104004} (\bibinfo {year} {2018})}\BibitemShut {NoStop}%
\bibitem [{\citenamefont {Ancilotto}\ \emph {et~al.}(2018)\citenamefont
  {Ancilotto}, \citenamefont {Barranco}, \citenamefont {Guilleumas},\ and\
  \citenamefont {Pi}}]{2018Ancilotto}%
  \BibitemOpen
  \bibfield  {author} {\bibinfo {author} {\bibfnamefont {F.}~\bibnamefont
  {Ancilotto}}, \bibinfo {author} {\bibfnamefont {M.}~\bibnamefont {Barranco}},
  \bibinfo {author} {\bibfnamefont {M.}~\bibnamefont {Guilleumas}}, \ and\
  \bibinfo {author} {\bibfnamefont {M.}~\bibnamefont {Pi}},\ }\href {\doibase
  10.1103/PhysRevA.98.053623} {\bibfield  {journal} {\bibinfo  {journal} {Phys.
  Rev. A}\ }\textbf {\bibinfo {volume} {98}},\ \bibinfo {pages} {053623}
  (\bibinfo {year} {2018})}\BibitemShut {NoStop}%
\bibitem [{\citenamefont {Tengstrand}\ \emph {et~al.}(2019)\citenamefont
  {Tengstrand}, \citenamefont {St\"urmer}, \citenamefont {Karabulut},\ and\
  \citenamefont {Reimann}}]{2019Tengstrand}%
  \BibitemOpen
  \bibfield  {author} {\bibinfo {author} {\bibfnamefont {M.~N.}\ \bibnamefont
  {Tengstrand}}, \bibinfo {author} {\bibfnamefont {P.}~\bibnamefont
  {St\"urmer}}, \bibinfo {author} {\bibfnamefont {E.~O.}\ \bibnamefont
  {Karabulut}}, \ and\ \bibinfo {author} {\bibfnamefont {S.~M.}\ \bibnamefont
  {Reimann}},\ }\href {\doibase 10.1103/PhysRevLett.123.160405} {\bibfield
  {journal} {\bibinfo  {journal} {Phys. Rev. Lett.}\ }\textbf {\bibinfo
  {volume} {123}},\ \bibinfo {pages} {160405} (\bibinfo {year}
  {2019})}\BibitemShut {NoStop}%
\bibitem [{\citenamefont {Gautam}\ and\ \citenamefont
  {Adhikari}(2019)}]{2019Gautam}%
  \BibitemOpen
  \bibfield  {author} {\bibinfo {author} {\bibfnamefont {S.}~\bibnamefont
  {Gautam}}\ and\ \bibinfo {author} {\bibfnamefont {S.~K.}\ \bibnamefont
  {Adhikari}},\ }\href@noop {} {\bibfield  {journal} {\bibinfo  {journal} {J.
  Phys. B}\ }\textbf {\bibinfo {volume} {52}},\ \bibinfo {pages} {055302}
  (\bibinfo {year} {2019})}\BibitemShut {NoStop}%
\bibitem [{\citenamefont {Cikojevi\ifmmode~\acute{c}\else \'{c}\fi{}}\ \emph
  {et~al.}(2019)\citenamefont {Cikojevi\ifmmode~\acute{c}\else \'{c}\fi{}},
  \citenamefont {Marki\ifmmode~\acute{c}\else \'{c}\fi{}}, \citenamefont
  {Astrakharchik},\ and\ \citenamefont {Boronat}}]{2019Cikojevi}%
  \BibitemOpen
  \bibfield  {author} {\bibinfo {author} {\bibfnamefont {V.}~\bibnamefont
  {Cikojevi\ifmmode~\acute{c}\else \'{c}\fi{}}}, \bibinfo {author}
  {\bibfnamefont {L.~V. c.~v.}\ \bibnamefont {Marki\ifmmode~\acute{c}\else
  \'{c}\fi{}}}, \bibinfo {author} {\bibfnamefont {G.~E.}\ \bibnamefont
  {Astrakharchik}}, \ and\ \bibinfo {author} {\bibfnamefont {J.}~\bibnamefont
  {Boronat}},\ }\href {\doibase 10.1103/PhysRevA.99.023618} {\bibfield
  {journal} {\bibinfo  {journal} {Phys. Rev. A}\ }\textbf {\bibinfo {volume}
  {99}},\ \bibinfo {pages} {023618} (\bibinfo {year} {2019})}\BibitemShut
  {NoStop}%
\bibitem [{\citenamefont {Kartashov}\ \emph {et~al.}(2019)\citenamefont
  {Kartashov}, \citenamefont {Astrakharchik}, \citenamefont {Malomed},\ and\
  \citenamefont {Torner}}]{2019kartashov}%
  \BibitemOpen
  \bibfield  {author} {\bibinfo {author} {\bibfnamefont {Y.~V.}\ \bibnamefont
  {Kartashov}}, \bibinfo {author} {\bibfnamefont {G.~E.}\ \bibnamefont
  {Astrakharchik}}, \bibinfo {author} {\bibfnamefont {B.~A.}\ \bibnamefont
  {Malomed}}, \ and\ \bibinfo {author} {\bibfnamefont {L.}~\bibnamefont
  {Torner}},\ }\href@noop {} {\bibfield  {journal} {\bibinfo  {journal} {Nat.
  Rev. Phys}\ }\textbf {\bibinfo {volume} {1}},\ \bibinfo {pages} {185}
  (\bibinfo {year} {2019})}\BibitemShut {NoStop}%
\bibitem [{\citenamefont {Li}\ \emph {et~al.}(2019)\citenamefont {Li},
  \citenamefont {Pan},\ and\ \citenamefont {Liu}}]{2019Li}%
  \BibitemOpen
  \bibfield  {author} {\bibinfo {author} {\bibfnamefont {Z.}~\bibnamefont
  {Li}}, \bibinfo {author} {\bibfnamefont {J.-S.}\ \bibnamefont {Pan}}, \ and\
  \bibinfo {author} {\bibfnamefont {W.~V.}\ \bibnamefont {Liu}},\ }\href
  {\doibase 10.1103/PhysRevA.100.053620} {\bibfield  {journal} {\bibinfo
  {journal} {Phys. Rev. A}\ }\textbf {\bibinfo {volume} {100}},\ \bibinfo
  {pages} {053620} (\bibinfo {year} {2019})}\BibitemShut {NoStop}%
\bibitem [{\citenamefont {Bulgac}(2002)}]{2002Bulgac}%
  \BibitemOpen
  \bibfield  {author} {\bibinfo {author} {\bibfnamefont {A.}~\bibnamefont
  {Bulgac}},\ }\href {\doibase 10.1103/PhysRevLett.89.050402} {\bibfield
  {journal} {\bibinfo  {journal} {Phys. Rev. Lett.}\ }\textbf {\bibinfo
  {volume} {89}},\ \bibinfo {pages} {050402} (\bibinfo {year}
  {2002})}\BibitemShut {NoStop}%
\bibitem [{\citenamefont {Petrov}(2015)}]{2015Petrov}%
  \BibitemOpen
  \bibfield  {author} {\bibinfo {author} {\bibfnamefont {D.~S.}\ \bibnamefont
  {Petrov}},\ }\href {\doibase 10.1103/PhysRevLett.115.155302} {\bibfield
  {journal} {\bibinfo  {journal} {Phys. Rev. Lett.}\ }\textbf {\bibinfo
  {volume} {115}},\ \bibinfo {pages} {155302} (\bibinfo {year}
  {2015})}\BibitemShut {NoStop}%
\bibitem [{\citenamefont {He}\ \emph {et~al.}(2019)\citenamefont {He},
  \citenamefont {Gao},\ and\ \citenamefont {Yu}}]{2019He}%
  \BibitemOpen
  \bibfield  {author} {\bibinfo {author} {\bibfnamefont {L.}~\bibnamefont
  {He}}, \bibinfo {author} {\bibfnamefont {P.}~\bibnamefont {Gao}}, \ and\
  \bibinfo {author} {\bibfnamefont {Z.-Q.}\ \bibnamefont {Yu}},\ }\href@noop {}
  {\bibfield  {journal} {\bibinfo  {journal} {arXiv:1910.12776}\ } (\bibinfo
  {year} {2019})}\BibitemShut {NoStop}%
\bibitem [{\citenamefont {Lee}\ \emph {et~al.}(1957)\citenamefont {Lee},
  \citenamefont {Huang},\ and\ \citenamefont {Yang}}]{1957Lee}%
  \BibitemOpen
  \bibfield  {author} {\bibinfo {author} {\bibfnamefont {T.~D.}\ \bibnamefont
  {Lee}}, \bibinfo {author} {\bibfnamefont {K.}~\bibnamefont {Huang}}, \ and\
  \bibinfo {author} {\bibfnamefont {C.~N.}\ \bibnamefont {Yang}},\ }\href
  {\doibase 10.1103/PhysRev.106.1135} {\bibfield  {journal} {\bibinfo
  {journal} {Phys. Rev.}\ }\textbf {\bibinfo {volume} {106}},\ \bibinfo {pages}
  {1135} (\bibinfo {year} {1957})}\BibitemShut {NoStop}%
\bibitem [{\citenamefont {Baillie}\ \emph {et~al.}(2016)\citenamefont
  {Baillie}, \citenamefont {Wilson}, \citenamefont {Bisset},\ and\
  \citenamefont {Blakie}}]{2016Baillie}%
  \BibitemOpen
  \bibfield  {author} {\bibinfo {author} {\bibfnamefont {D.}~\bibnamefont
  {Baillie}}, \bibinfo {author} {\bibfnamefont {R.~M.}\ \bibnamefont {Wilson}},
  \bibinfo {author} {\bibfnamefont {R.~N.}\ \bibnamefont {Bisset}}, \ and\
  \bibinfo {author} {\bibfnamefont {P.~B.}\ \bibnamefont {Blakie}},\ }\href
  {\doibase 10.1103/PhysRevA.94.021602} {\bibfield  {journal} {\bibinfo
  {journal} {Phys. Rev. A}\ }\textbf {\bibinfo {volume} {94}},\ \bibinfo
  {pages} {021602} (\bibinfo {year} {2016})}\BibitemShut {NoStop}%
\bibitem [{\citenamefont {Petrov}\ and\ \citenamefont
  {Astrakharchik}(2016)}]{2016Petrov}%
  \BibitemOpen
  \bibfield  {author} {\bibinfo {author} {\bibfnamefont {D.~S.}\ \bibnamefont
  {Petrov}}\ and\ \bibinfo {author} {\bibfnamefont {G.~E.}\ \bibnamefont
  {Astrakharchik}},\ }\href {\doibase 10.1103/PhysRevLett.117.100401}
  {\bibfield  {journal} {\bibinfo  {journal} {Phys. Rev. Lett.}\ }\textbf
  {\bibinfo {volume} {117}},\ \bibinfo {pages} {100401} (\bibinfo {year}
  {2016})}\BibitemShut {NoStop}%
\bibitem [{\citenamefont {Sekino}\ and\ \citenamefont
  {Nishida}(2018)}]{2018Sekino}%
  \BibitemOpen
  \bibfield  {author} {\bibinfo {author} {\bibfnamefont {Y.}~\bibnamefont
  {Sekino}}\ and\ \bibinfo {author} {\bibfnamefont {Y.}~\bibnamefont
  {Nishida}},\ }\href {\doibase 10.1103/PhysRevA.97.011602} {\bibfield
  {journal} {\bibinfo  {journal} {Phys. Rev. A}\ }\textbf {\bibinfo {volume}
  {97}},\ \bibinfo {pages} {011602} (\bibinfo {year} {2018})}\BibitemShut
  {NoStop}%
\bibitem [{\citenamefont {Ilg}\ \emph {et~al.}(2018)\citenamefont {Ilg},
  \citenamefont {Kumlin}, \citenamefont {Santos}, \citenamefont {Petrov},\ and\
  \citenamefont {B\"uchler}}]{2018Petrov}%
  \BibitemOpen
  \bibfield  {author} {\bibinfo {author} {\bibfnamefont {T.}~\bibnamefont
  {Ilg}}, \bibinfo {author} {\bibfnamefont {J.}~\bibnamefont {Kumlin}},
  \bibinfo {author} {\bibfnamefont {L.}~\bibnamefont {Santos}}, \bibinfo
  {author} {\bibfnamefont {D.~S.}\ \bibnamefont {Petrov}}, \ and\ \bibinfo
  {author} {\bibfnamefont {H.~P.}\ \bibnamefont {B\"uchler}},\ }\href {\doibase
  10.1103/PhysRevA.98.051604} {\bibfield  {journal} {\bibinfo  {journal} {Phys.
  Rev. A}\ }\textbf {\bibinfo {volume} {98}},\ \bibinfo {pages} {051604}
  (\bibinfo {year} {2018})}\BibitemShut {NoStop}%
\bibitem [{\citenamefont {Li}\ \emph {et~al.}(2018)\citenamefont {Li},
  \citenamefont {Chen}, \citenamefont {Luo}, \citenamefont {Huang},
  \citenamefont {Tan}, \citenamefont {Pang},\ and\ \citenamefont
  {Malomed}}]{2018Li}%
  \BibitemOpen
  \bibfield  {author} {\bibinfo {author} {\bibfnamefont {Y.}~\bibnamefont
  {Li}}, \bibinfo {author} {\bibfnamefont {Z.}~\bibnamefont {Chen}}, \bibinfo
  {author} {\bibfnamefont {Z.}~\bibnamefont {Luo}}, \bibinfo {author}
  {\bibfnamefont {C.}~\bibnamefont {Huang}}, \bibinfo {author} {\bibfnamefont
  {H.}~\bibnamefont {Tan}}, \bibinfo {author} {\bibfnamefont {W.}~\bibnamefont
  {Pang}}, \ and\ \bibinfo {author} {\bibfnamefont {B.~A.}\ \bibnamefont
  {Malomed}},\ }\href {\doibase 10.1103/PhysRevA.98.063602} {\bibfield
  {journal} {\bibinfo  {journal} {Phys. Rev. A}\ }\textbf {\bibinfo {volume}
  {98}},\ \bibinfo {pages} {063602} (\bibinfo {year} {2018})}\BibitemShut
  {NoStop}%
\bibitem [{\citenamefont {Chiquillo}(2019)}]{2019Chiquillo}%
  \BibitemOpen
  \bibfield  {author} {\bibinfo {author} {\bibfnamefont {E.}~\bibnamefont
  {Chiquillo}},\ }\href {\doibase 10.1103/PhysRevA.99.051601} {\bibfield
  {journal} {\bibinfo  {journal} {Phys. Rev. A}\ }\textbf {\bibinfo {volume}
  {99}},\ \bibinfo {pages} {051601} (\bibinfo {year} {2019})}\BibitemShut
  {NoStop}%
\bibitem [{\citenamefont {Aybar}\ and\ \citenamefont
  {Oktel}(2019)}]{2019Aybar}%
  \BibitemOpen
  \bibfield  {author} {\bibinfo {author} {\bibfnamefont {E.}~\bibnamefont
  {Aybar}}\ and\ \bibinfo {author} {\bibfnamefont {M.~O.}\ \bibnamefont
  {Oktel}},\ }\href {\doibase 10.1103/PhysRevA.99.013620} {\bibfield  {journal}
  {\bibinfo  {journal} {Phys. Rev. A}\ }\textbf {\bibinfo {volume} {99}},\
  \bibinfo {pages} {013620} (\bibinfo {year} {2019})}\BibitemShut {NoStop}%
\bibitem [{\citenamefont {Wilson}\ \emph {et~al.}(2018)\citenamefont {Wilson},
  \citenamefont {Westerberg}, \citenamefont {Valiente}, \citenamefont {Duncan},
  \citenamefont {Wright}, \citenamefont {\"Ohberg},\ and\ \citenamefont
  {Faccio}}]{2018Wilson}%
  \BibitemOpen
  \bibfield  {author} {\bibinfo {author} {\bibfnamefont {K.~E.}\ \bibnamefont
  {Wilson}}, \bibinfo {author} {\bibfnamefont {N.}~\bibnamefont {Westerberg}},
  \bibinfo {author} {\bibfnamefont {M.}~\bibnamefont {Valiente}}, \bibinfo
  {author} {\bibfnamefont {C.~W.}\ \bibnamefont {Duncan}}, \bibinfo {author}
  {\bibfnamefont {E.~M.}\ \bibnamefont {Wright}}, \bibinfo {author}
  {\bibfnamefont {P.}~\bibnamefont {\"Ohberg}}, \ and\ \bibinfo {author}
  {\bibfnamefont {D.}~\bibnamefont {Faccio}},\ }\href {\doibase
  10.1103/PhysRevLett.121.133903} {\bibfield  {journal} {\bibinfo  {journal}
  {Phys. Rev. Lett.}\ }\textbf {\bibinfo {volume} {121}},\ \bibinfo {pages}
  {133903} (\bibinfo {year} {2018})}\BibitemShut {NoStop}%
\bibitem [{\citenamefont {Westerberg}\ \emph {et~al.}(2018)\citenamefont
  {Westerberg}, \citenamefont {Wilson}, \citenamefont {Duncan}, \citenamefont
  {Faccio}, \citenamefont {Wright}, \citenamefont {\"Ohberg},\ and\
  \citenamefont {Valiente}}]{2018Westerberg}%
  \BibitemOpen
  \bibfield  {author} {\bibinfo {author} {\bibfnamefont {N.}~\bibnamefont
  {Westerberg}}, \bibinfo {author} {\bibfnamefont {K.~E.}\ \bibnamefont
  {Wilson}}, \bibinfo {author} {\bibfnamefont {C.~W.}\ \bibnamefont {Duncan}},
  \bibinfo {author} {\bibfnamefont {D.}~\bibnamefont {Faccio}}, \bibinfo
  {author} {\bibfnamefont {E.~M.}\ \bibnamefont {Wright}}, \bibinfo {author}
  {\bibfnamefont {P.}~\bibnamefont {\"Ohberg}}, \ and\ \bibinfo {author}
  {\bibfnamefont {M.}~\bibnamefont {Valiente}},\ }\href {\doibase
  10.1103/PhysRevA.98.053835} {\bibfield  {journal} {\bibinfo  {journal} {Phys.
  Rev. A}\ }\textbf {\bibinfo {volume} {98}},\ \bibinfo {pages} {053835}
  (\bibinfo {year} {2018})}\BibitemShut {NoStop}%
\bibitem [{\citenamefont {Salasnich}\ \emph {et~al.}(2007)\citenamefont
  {Salasnich}, \citenamefont {Adhikari},\ and\ \citenamefont
  {Toigo}}]{2007Salasnich}%
  \BibitemOpen
  \bibfield  {author} {\bibinfo {author} {\bibfnamefont {L.}~\bibnamefont
  {Salasnich}}, \bibinfo {author} {\bibfnamefont {S.~K.}\ \bibnamefont
  {Adhikari}}, \ and\ \bibinfo {author} {\bibfnamefont {F.}~\bibnamefont
  {Toigo}},\ }\href {\doibase 10.1103/PhysRevA.75.023616} {\bibfield  {journal}
  {\bibinfo  {journal} {Phys. Rev. A}\ }\textbf {\bibinfo {volume} {75}},\
  \bibinfo {pages} {023616} (\bibinfo {year} {2007})}\BibitemShut {NoStop}%
\bibitem [{\citenamefont {Cui}(2018)}]{2018Cui}%
  \BibitemOpen
  \bibfield  {author} {\bibinfo {author} {\bibfnamefont {X.}~\bibnamefont
  {Cui}},\ }\href {\doibase 10.1103/PhysRevA.98.023630} {\bibfield  {journal}
  {\bibinfo  {journal} {Phys. Rev. A}\ }\textbf {\bibinfo {volume} {98}},\
  \bibinfo {pages} {023630} (\bibinfo {year} {2018})}\BibitemShut {NoStop}%
\bibitem [{\citenamefont {Adhikari}(2018)}]{2018Adhikari}%
  \BibitemOpen
  \bibfield  {author} {\bibinfo {author} {\bibfnamefont {S.}~\bibnamefont
  {Adhikari}},\ }\href@noop {} {\bibfield  {journal} {\bibinfo  {journal}
  {Laser Phys. Lett}\ }\textbf {\bibinfo {volume} {15}},\ \bibinfo {pages}
  {095501} (\bibinfo {year} {2018})}\BibitemShut {NoStop}%
\bibitem [{\citenamefont {Rakshit}\ \emph
  {et~al.}(2019{\natexlab{a}})\citenamefont {Rakshit}, \citenamefont {Karpiuk},
  \citenamefont {Brewczyk},\ and\ \citenamefont {Gajda}}]{2019Rakshit}%
  \BibitemOpen
  \bibfield  {author} {\bibinfo {author} {\bibfnamefont {D.}~\bibnamefont
  {Rakshit}}, \bibinfo {author} {\bibfnamefont {T.}~\bibnamefont {Karpiuk}},
  \bibinfo {author} {\bibfnamefont {M.}~\bibnamefont {Brewczyk}}, \ and\
  \bibinfo {author} {\bibfnamefont {M.}~\bibnamefont {Gajda}},\ }\href@noop {}
  {\bibfield  {journal} {\bibinfo  {journal} {SciPost Phys}\ }\textbf {\bibinfo
  {volume} {6}},\ \bibinfo {pages} {079} (\bibinfo {year}
  {2019}{\natexlab{a}})}\BibitemShut {NoStop}%
\bibitem [{\citenamefont {Rakshit}\ \emph
  {et~al.}(2019{\natexlab{b}})\citenamefont {Rakshit}, \citenamefont {Karpiuk},
  \citenamefont {Zin}, \citenamefont {Brewczyk}, \citenamefont {Lewenstein},\
  and\ \citenamefont {Gajda}}]{2019Rakshittwo}%
  \BibitemOpen
  \bibfield  {author} {\bibinfo {author} {\bibfnamefont {D.}~\bibnamefont
  {Rakshit}}, \bibinfo {author} {\bibfnamefont {T.}~\bibnamefont {Karpiuk}},
  \bibinfo {author} {\bibfnamefont {P.}~\bibnamefont {Zin}}, \bibinfo {author}
  {\bibfnamefont {M.}~\bibnamefont {Brewczyk}}, \bibinfo {author}
  {\bibfnamefont {M.}~\bibnamefont {Lewenstein}}, \ and\ \bibinfo {author}
  {\bibfnamefont {M.}~\bibnamefont {Gajda}},\ }\href {\doibase
  10.1088/1367-2630/ab2ce3} {\bibfield  {journal} {\bibinfo  {journal} {New J.
  Phys}\ }\textbf {\bibinfo {volume} {21}},\ \bibinfo {pages} {073027}
  (\bibinfo {year} {2019}{\natexlab{b}})}\BibitemShut {NoStop}%
\bibitem [{\citenamefont {Albus}\ \emph {et~al.}(2002)\citenamefont {Albus},
  \citenamefont {Gardiner}, \citenamefont {Illuminati},\ and\ \citenamefont
  {Wilkens}}]{2002Albus}%
  \BibitemOpen
  \bibfield  {author} {\bibinfo {author} {\bibfnamefont {A.~P.}\ \bibnamefont
  {Albus}}, \bibinfo {author} {\bibfnamefont {S.~A.}\ \bibnamefont {Gardiner}},
  \bibinfo {author} {\bibfnamefont {F.}~\bibnamefont {Illuminati}}, \ and\
  \bibinfo {author} {\bibfnamefont {M.}~\bibnamefont {Wilkens}},\ }\href
  {\doibase 10.1103/PhysRevA.65.053607} {\bibfield  {journal} {\bibinfo
  {journal} {Phys. Rev. A}\ }\textbf {\bibinfo {volume} {65}},\ \bibinfo
  {pages} {053607} (\bibinfo {year} {2002})}\BibitemShut {NoStop}%
\bibitem [{\citenamefont {Viverit}\ and\ \citenamefont
  {Giorgini}(2002)}]{2002Viverit}%
  \BibitemOpen
  \bibfield  {author} {\bibinfo {author} {\bibfnamefont {L.}~\bibnamefont
  {Viverit}}\ and\ \bibinfo {author} {\bibfnamefont {S.}~\bibnamefont
  {Giorgini}},\ }\href {\doibase 10.1103/PhysRevA.66.063604} {\bibfield
  {journal} {\bibinfo  {journal} {Phys. Rev. A}\ }\textbf {\bibinfo {volume}
  {66}},\ \bibinfo {pages} {063604} (\bibinfo {year} {2002})}\BibitemShut
  {NoStop}%
\bibitem [{\citenamefont {Ferrier-Barbut}\ \emph {et~al.}(2014)\citenamefont
  {Ferrier-Barbut}, \citenamefont {Delehaye}, \citenamefont {Laurent},
  \citenamefont {Grier}, \citenamefont {Pierce}, \citenamefont {Rem},
  \citenamefont {Chevy},\ and\ \citenamefont {Salomon}}]{2014Ferrier}%
  \BibitemOpen
  \bibfield  {author} {\bibinfo {author} {\bibfnamefont {I.}~\bibnamefont
  {Ferrier-Barbut}}, \bibinfo {author} {\bibfnamefont {M.}~\bibnamefont
  {Delehaye}}, \bibinfo {author} {\bibfnamefont {S.}~\bibnamefont {Laurent}},
  \bibinfo {author} {\bibfnamefont {A.~T.}\ \bibnamefont {Grier}}, \bibinfo
  {author} {\bibfnamefont {M.}~\bibnamefont {Pierce}}, \bibinfo {author}
  {\bibfnamefont {B.~S.}\ \bibnamefont {Rem}}, \bibinfo {author} {\bibfnamefont
  {F.}~\bibnamefont {Chevy}}, \ and\ \bibinfo {author} {\bibfnamefont
  {C.}~\bibnamefont {Salomon}},\ }\href@noop {} {\bibfield  {journal} {\bibinfo
   {journal} {Science}\ }\textbf {\bibinfo {volume} {345}},\ \bibinfo {pages}
  {1035} (\bibinfo {year} {2014})}\BibitemShut {NoStop}%
\bibitem [{\citenamefont {DeSalvo}\ \emph {et~al.}(2017)\citenamefont
  {DeSalvo}, \citenamefont {Patel}, \citenamefont {Johansen},\ and\
  \citenamefont {Chin}}]{2017DeSalvo}%
  \BibitemOpen
  \bibfield  {author} {\bibinfo {author} {\bibfnamefont {B.~J.}\ \bibnamefont
  {DeSalvo}}, \bibinfo {author} {\bibfnamefont {K.}~\bibnamefont {Patel}},
  \bibinfo {author} {\bibfnamefont {J.}~\bibnamefont {Johansen}}, \ and\
  \bibinfo {author} {\bibfnamefont {C.}~\bibnamefont {Chin}},\ }\href {\doibase
  10.1103/PhysRevLett.119.233401} {\bibfield  {journal} {\bibinfo  {journal}
  {Phys. Rev. Lett.}\ }\textbf {\bibinfo {volume} {119}},\ \bibinfo {pages}
  {233401} (\bibinfo {year} {2017})}\BibitemShut {NoStop}%
\bibitem [{\citenamefont {Bloch}\ \emph {et~al.}(2008)\citenamefont {Bloch},
  \citenamefont {Dalibard},\ and\ \citenamefont {Zwerger}}]{2008Bloch}%
  \BibitemOpen
  \bibfield  {author} {\bibinfo {author} {\bibfnamefont {I.}~\bibnamefont
  {Bloch}}, \bibinfo {author} {\bibfnamefont {J.}~\bibnamefont {Dalibard}}, \
  and\ \bibinfo {author} {\bibfnamefont {W.}~\bibnamefont {Zwerger}},\ }\href
  {\doibase 10.1103/RevModPhys.80.885} {\bibfield  {journal} {\bibinfo
  {journal} {Rev. Mod. Phys.}\ }\textbf {\bibinfo {volume} {80}},\ \bibinfo
  {pages} {885} (\bibinfo {year} {2008})}\BibitemShut {NoStop}%
\bibitem [{\citenamefont {Zhang}\ \emph {et~al.}(2014)\citenamefont {Zhang},
  \citenamefont {Zhang}, \citenamefont {Zhai},\ and\ \citenamefont
  {Zhang}}]{2014Zhang}%
  \BibitemOpen
  \bibfield  {author} {\bibinfo {author} {\bibfnamefont {R.}~\bibnamefont
  {Zhang}}, \bibinfo {author} {\bibfnamefont {W.}~\bibnamefont {Zhang}},
  \bibinfo {author} {\bibfnamefont {H.}~\bibnamefont {Zhai}}, \ and\ \bibinfo
  {author} {\bibfnamefont {P.}~\bibnamefont {Zhang}},\ }\href {\doibase
  10.1103/PhysRevA.90.063614} {\bibfield  {journal} {\bibinfo  {journal} {Phys.
  Rev. A}\ }\textbf {\bibinfo {volume} {90}},\ \bibinfo {pages} {063614}
  (\bibinfo {year} {2014})}\BibitemShut {NoStop}%
\bibitem [{\citenamefont {Cui}(2014)}]{2014Cui}%
  \BibitemOpen
  \bibfield  {author} {\bibinfo {author} {\bibfnamefont {X.}~\bibnamefont
  {Cui}},\ }\href {\doibase 10.1103/PhysRevA.90.041603} {\bibfield  {journal}
  {\bibinfo  {journal} {Phys. Rev. A}\ }\textbf {\bibinfo {volume} {90}},\
  \bibinfo {pages} {041603} (\bibinfo {year} {2014})}\BibitemShut {NoStop}%
\bibitem [{\citenamefont {Petrov}\ \emph {et~al.}(2004)\citenamefont {Petrov},
  \citenamefont {Salomon},\ and\ \citenamefont {Shlyapnikov}}]{2004Petrov}%
  \BibitemOpen
  \bibfield  {author} {\bibinfo {author} {\bibfnamefont {D.~S.}\ \bibnamefont
  {Petrov}}, \bibinfo {author} {\bibfnamefont {C.}~\bibnamefont {Salomon}}, \
  and\ \bibinfo {author} {\bibfnamefont {G.~V.}\ \bibnamefont {Shlyapnikov}},\
  }\href {\doibase 10.1103/PhysRevLett.93.090404} {\bibfield  {journal}
  {\bibinfo  {journal} {Phys. Rev. Lett.}\ }\textbf {\bibinfo {volume} {93}},\
  \bibinfo {pages} {090404} (\bibinfo {year} {2004})}\BibitemShut {NoStop}%
\bibitem [{\citenamefont {Yi}\ and\ \citenamefont {Cui}(2015)}]{2015Yi}%
  \BibitemOpen
  \bibfield  {author} {\bibinfo {author} {\bibfnamefont {W.}~\bibnamefont
  {Yi}}\ and\ \bibinfo {author} {\bibfnamefont {X.}~\bibnamefont {Cui}},\
  }\href {\doibase 10.1103/PhysRevA.92.013620} {\bibfield  {journal} {\bibinfo
  {journal} {Phys. Rev. A}\ }\textbf {\bibinfo {volume} {92}},\ \bibinfo
  {pages} {013620} (\bibinfo {year} {2015})}\BibitemShut {NoStop}%
\end{thebibliography}%

\end{document}